# Enabling Loosely-Coupled Serial Job Execution on the IBM BlueGene/P Supercomputer and the SiCortex SC5832

Ioan Raicu[*], Zhao Zhang[+], Mike Wilde[#+], Ian Foster[#*+]
[*]Department of Computer Science, University of Chicago, IL, USA
[+]Computation Institute, University of Chicago & Argonne National Laboratory, USA
[#]Math and Computer Science Division, Argonne National Laboratory, Argonne IL, USA
iraicu@cs.uchicago.edu, zhaozhang@uchicago.edu, wilde@mcs.anl.gov, foster@mcs.anl.gov

## Abstract

*Our work addresses the enabling of the execution of highly parallel computations composed of loosely coupled serial jobs with no modifications to the respective applications, on large-scale systems. This approach allows new-and potentially far larger-classes of application to leverage systems such as the IBM Blue Gene/P supercomputer and similar emerging petascale architectures. We present here the challenges of I/O performance encountered in making this model practical, and show results using both micro-benchmarks and real applications on two large-scale systems, the BG/P and the SiCortex SC5832. Our preliminary benchmarks show that we can scale to 4096 processors on the Blue Gene/P and 5832 processors on the SiCortex with high efficiency, and can achieve thousands of tasks/sec sustained execution rates for parallel workloads of ordinary serial applications. We measured applications from two domains, economic energy modeling and molecular dynamics.*

## 1. Introduction

Emerging petascale computing systems are primarily dedicated to tightly coupled, massively parallel applications implemented using message passing paradigms. Such systems—typified by IBM's Blue Gene/P [1]—include fast integrated custom interconnects, multi-core processors, and multi-level I/O subsystems, technologies that are also found in smaller, lower-cost, and energy-efficient systems such as the SiCortex SC5832 [2]. These architectures are well suited for a large class of applications that require a tightly coupled programming approach. However, there is a potentially larger class of "ordinary" serial applications that are precluded from leveraging the increasing power of modern parallel systems due to the lack of efficient support in those systems for the "scripting" programming model in which application and utility programs are linked into useful workflows through the looser task-coupling model of passing data via files.

With the advances in e-Sciences and the growing complexity of scientific analyses, more and more scientists and researchers are relying on various forms of application scripting systems to automate the workflow of process coordination, derivation automation, provenance tracking, and bookkeeping. Their approaches are typically based on a model of loosely coupled computation, exchanging data via files, databases or XML documents, or a combination of these. Furthermore, with technology advances in both scientific instrumentation and simulation, the volume of scientific datasets is growing exponentially. This vast increase in data volume combined with the growing complexity of data analysis procedures and algorithms have rendered traditional manual and even automated serial processing and exploration unfavorable as compared with modern high performance computing processes automated by scientific workflow systems.

We focus in this paper on the ability to execute large scale applications leveraging existing scripting systems on petascale systems such as the IBM Blue Gene/P. Blue Gene-class systems have been traditionally called high performance computing (HPC) systems, as they almost exclusively execute tightly coupled parallel jobs within a particular machine over low-latency interconnects; the applications typically use a message passing interface (e.g. MPI) to achieve the needed inter-process communication. Conversely, high throughput computing (HTC) systems (which scientific workflows can more readily utilize) generally involve the execution of independent, sequential jobs that can be individually scheduled on many different computing resources across multiple administrative boundaries. HTC systems achieve this using various grid computing techniques, and almost exclusively use files, documents or databases rather than messages for inter-process communication.

The hypothesis is that loosely coupled applications can be executed efficiently on today's supercomputers; this paper provides empirical evidence to prove our hypothesis. The paper also describes the set of problems that must be overcome to make loosely-coupled programming practical on emerging petascale architectures: local resource manager scalability and granularity, efficient utilization of the raw hardware, shared file system contention, and application scalability. It describes how we address these problems, and identifies the remaining problems that need to be solved to make loosely coupled supercomputing a practical and accepted reality. Through our work, we have enabled the BG/P to efficiently support loosely coupled parallel programming without any modifications to the respective applications, enabling the same applications that execute in a distributed grid environment to be run efficiently on the BG/P and similar systems.

We validate our hypothesis by testing and measuring two systems, Swift and Falkon, which have been used extensively to execute large-scale loosely coupled applications on clusters and grids. We present results of both micro-benchmarks and real applications executed on two representative large scale systems, the BG/P and the SiCortex SC5832. Focused micro-benchmarks show that we can scale to thousands of processors with high efficiency, and can achieve sustained execution rates of thousands of tasks per second. We investigate two applications from different domains, economic energy modeling and molecular dynamics, and show excellent speedup and efficiency as they scale to thousands of processors.

## 2. Related Work

Due to the only recent availability of parallel systems with 10K cores or more, and the even scarcer experience or success in loosely-coupled programming at this scale, we find that there is little existing work with which we can compare. The Condor high-throughput system, and in particular its concept of glide-ins, has been compared to Falkon in previous papers [3]. This system was evaluated on the BG/L system [4] is currently being tested on the BG/P architecture, but performance measurements and application experiences are not yet published. Other local resource managers were evaluated in [3] as well.

In the world of high throughput computing, systems such as Map-Reduce [5], Hadoop [6] and BOINC [7] have utilized highly distributed pools of processors, but the focus (and metrics) of these systems has not been on single highly-parallel machines such as those we focus on here. Map/reduce is typically applied to a data model consisting of name/value pairs, processed at the programming language level. It has several similarities to the approach we apply here, in particular its ability to spread the processing of a large dataset to thousands of processors. However, it is far less amenable to the utilization and chaining of exiting application programs, and often involves the development of custom filtering scripts. An approach by Reid called "task farming" [8], also at the programming language level, has been evaluated on the BG/L.

Coordination languages developed in the 1980s and 1990s [9, 10, 11, 12], describe individual computation components and their ports and channels, and the data and event flow between them. They also coordinate the execution of the components, often on parallel computing resources. In the scientific community there are a number of emerging systems for scientific programming and computation [15, 5, 13, 14 Our work here builds on the Swift parallel programming system [16, 15], in large part because its programming model abstracts the unit of data passing as a dataset rather than directly exposing sets of files, and because its data-flow model exposes the information needed to efficiently compile it for a wide variety of architectures while maintaining a simple common programming model. Detailed data flow analysis towards automating the compilation of conventional in-memory programming models for thousands of cores is described by Hwu et. al [40], but this does not address the simpler and more accessible data flow approach taken by Swift for scripting loosely coupled applications, which leverages implicit parallelism as well.

## 3. Requirements and Implementation

Our goal of running loosely-coupled applications efficiently and routinely on petascale systems requires that both HTC and HPC applications be able to co-exist on systems such as the BG/P. Petascale systems have been designed as HPC systems, so it is not surprising that the naïve use of these systems for HTC applications yields poor utilization and performance. The contribution of the work we describe here is the ability to enable a new class of applications – large-scale loosely-coupled – to efficiently execute on petascale systems. This is accomplished primarily through three mechanisms: 1) multi-level scheduling, 2) efficient task dispatch, and 3) extensive use of caching to avoid shared infrastructure (e.g. file systems and interconnects).

Multi-level scheduling is essential on a system such as the BG/P because the local resource manager (LRM, Cobalt [17]) works at a granularity of processor-sets, or PSETs [18], rather than individual computing nodes or processor cores. On the BG/P, a PSET is a group of 64 compute nodes (each with 4 processor cores) and one I/O node. PSETs must be allocated in their entirety to user application jobs by the LRM, which imposes the constraint that the applications must make use of all 256 cores, or waste valuable CPU resources. Tightly coupled MPI applications are well suited for this constraint, but loosely-coupled application workflows, on the other hand, generally have many single processor jobs, each with possibly unique executables and almost always with unique parameters. Naively running such applications on the BG/P using the system's Cobalt LRM would yield, at worst case, a 1/256 utilization if the single processor job is not multi-threaded, or 1/64 if it is. In the work we describe here, we use multi-level scheduling to allocate compute resources from Cobalt at the PSET granularity, and then make these computational resources available to applications at a single processor core granularity in order to enable single threaded jobs to execute with up to 100% utilization. Using this multi-level scheduling mechanism, we are able to launch a unique application, or the same application with unique arguments, on each core, and to launch such tasks repetitively throughout the allocation period. This capability is made possible through Falkon and its resource provisioning mechanisms.

A related obstacle to loosely coupled programming when using the native BG/P LRM is the overhead of scheduling and starting resources. The BG/P compute nodes are powered off when not in use, and hence must be booted when allocated to a job. As the compute nodes do not have local disks, the boot up process involves reading a Linux (or POSIX-based ZeptOS [19]) kernel image from a shared file system, which can be expensive if compute nodes are allocated and de-allocated frequently. Using multi-level scheduling allows this high initial cost to be amortized over many jobs, reducing it to an insignificant overhead. With the use of multi-level scheduling, executing a job is reduced to its bare and lightweight essentials: loading the application into memory, executing it, and returning its exit code – a process that can occur in milliseconds. We contrast this with the cost to reboot compute nodes, which is on the order of multiple seconds (for a single node) and can be as high as hundreds of seconds if many compute nodes are rebooting concurrently.

The second mechanism that enables loosely-coupled applications to be executed on the BG/P is a streamlined task submission framework (Falkon [3]) which provides efficient task dispatch. This is made possible through Falkon's focus: it relies on LRMs for many functions (e.g., reservation, policy-based scheduling, accounting) and client frameworks such as workflow systems or distributed scripting systems for others (e.g., recovery, data staging, job dependency management). In contrast with typical LRM performance, Falkon achieves several orders of magnitude higher performance (607 to 3773 tasks/sec in a Linux cluster environment, 3057 tasks/sec on the SiCortex, 1758 tasks/sec on BG/P).

To quantify the need and benefit of such high throughputs, we analyze (in Figure 1 and Figure 2) the achievable resource efficiency at two different supercomputer scales (the 4K processors of the BG/P test system currently available to us, and the ultimate size of the ALCF BG/P system – 160K processors [1, 20]) for various throughput levels (1, 10, 100, 1K, and 10K tasks/sec). This graph shows the minimum task durations needed to achieve a range of efficiencies up to 1.0 (100%), given the peak task submission rates of the various actual and hypothetical task scheduling facilities. In this analysis, efficiency is defined as *(achieved speedup) / (ideal speedup)*.

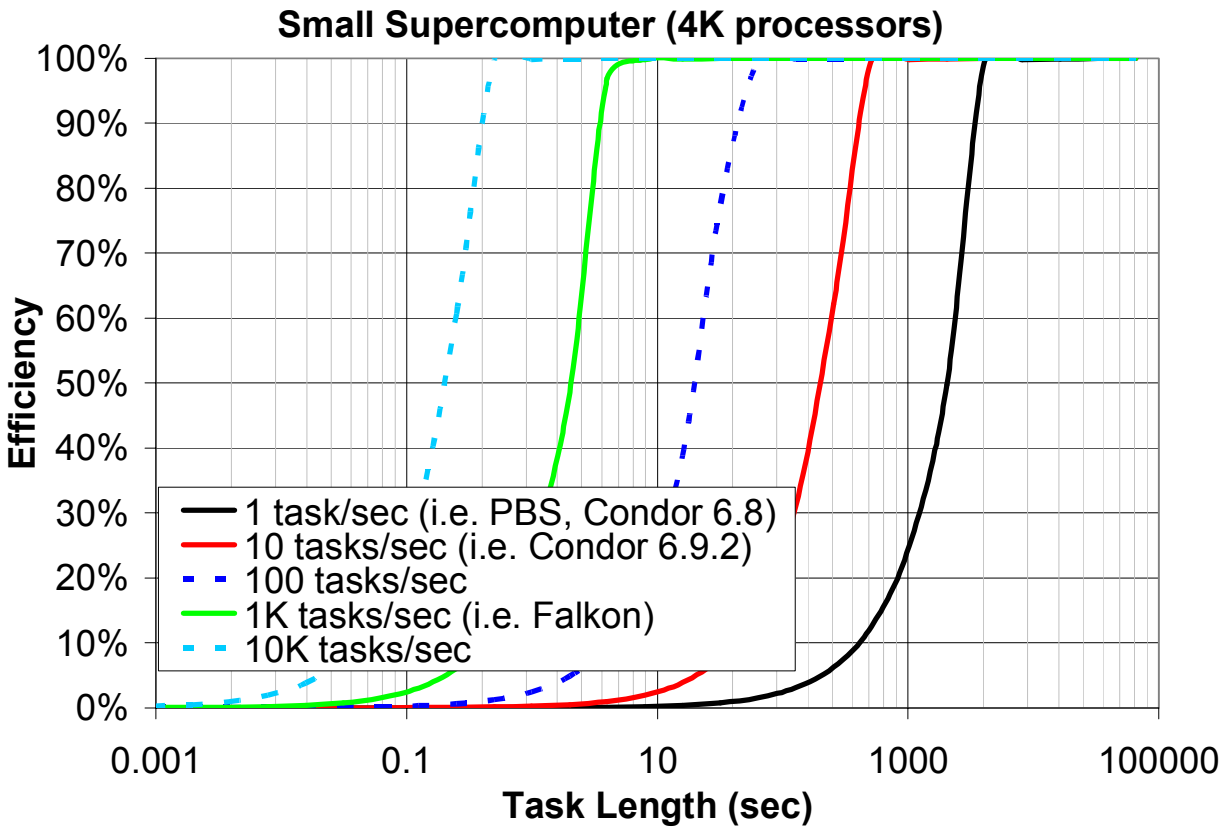

**Figure 1: Theoretical resource efficiency for both a small and large supercomputer in executing 1M tasks of various lengths at various dispatch rates**

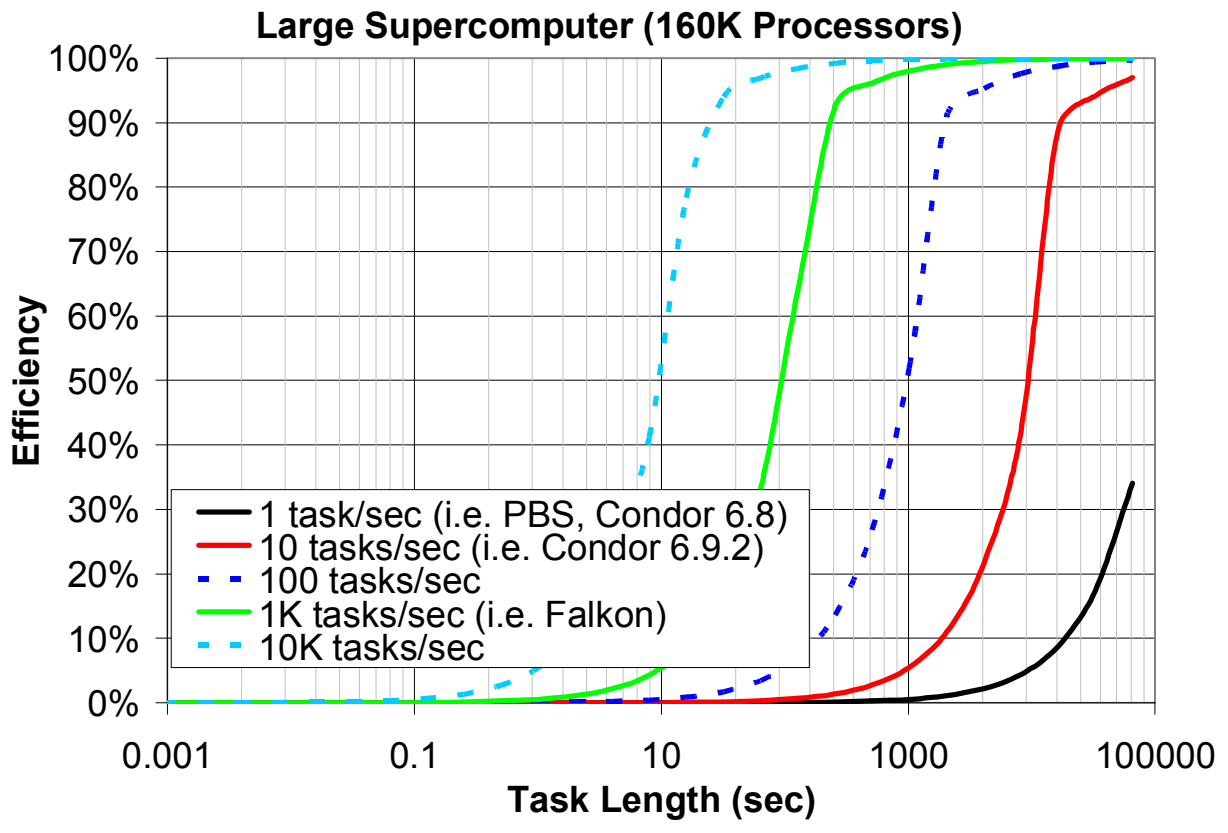

**Figure 2: Theoretical resource efficiency for both a small and large supercomputer in executing 1M tasks of various lengths at various dispatch rates**

We note that current production LRMs require relatively long tasks in order to maintain high efficiency. For example, in the small supercomputer case with 4096 processors, for a scheduler that can submit 10 tasks/sec, tasks need to be 520 seconds in duration in order to get even 90% efficiency; for the same throughput, the required task duration is increased to 30,000 seconds (~9 hours) for 160K processors in order to maintain 90% efficiency. With throughputs of 1000 tasks/sec (which Falkon can sustain on the BG/P), the same 90% efficiency can be reached with tasks of length 3.75 seconds and 256 seconds for the same two cases. Empirical results presented in a later section (Figure 8) shows that on the BG/P with 2048 processors, 4 second tasks yield 94% efficiency, and on the SiCortex with 5760 processors, we need 8 second tasks to achieve the same 94% efficiency. The calculations in this analysis show that the higher the throughput rates (tasks/sec) that can be dispatched to and executed on a set of resources, the higher the maximum resource utilization and efficiency for the same workloads and the faster the applications turn-around times will be, assuming the I/O needed for the application scales with the number of processors.

Finally, the third mechanism we employ for enabling loosely coupled applications to execute efficiently on the BG/P is extensive use of caching to allow better application scalability by avoiding shared file systems. As workflow systems frequently employ files as the primary communication medium between data-dependent jobs, having efficient mechanisms to read and write files is critical. The compute nodes on the BG/P do not have local disks, but they have both a shared file system (GPFS [21]) and local filesystem implemented in RAM (“ramdisk”); the SiCortex is similar as it also has a shared file system (NFS and PVFS), as well as a local file system in RAM. We make extensive use of the ramdisk local filesystem, to cache file objects such as application scripts and binary executables, static input data which is constant across many jobs running an application, and in some cases, output data from the application until enough data is collected to allow efficient writes to the shared file system. We found that naively executing applications directly against the shared file system yielded unacceptably poor performance, but with successive levels of caching we were able to increase the execution efficiency to within a few percent of ideal in many cases.

## 3.1 Swift and Falkon

To harness a wide array of loosely coupled applications that have already been implemented and executed in clusters and grids, we decided to build upon existing systems (Swift [15] and Falkon [3]). Swift is an emerging system that enables scientific workflows through a data-flow-based functional parallel programming model. It is a parallel scripting tool for rapid and reliable specification, execution, and management of large-scale science and engineering workflows. Swift takes a structured approach to workflow specification, scheduling and execution. It consists of a simple functional scripting language called SwiftScript for concise specifications of complex parallel computations based on dataset typing and iterators, and dynamic dataset mappings for accessing complex, large scale datasets represented in diverse data formats.

The runtime system in Swift relies on the CoG Karajan [22] workflow engine for efficient scheduling and load balancing, and it integrates with the Falkon [3] light-weight task execution service for optimized task throughput and resource efficiency. Falkon enables the rapid and efficient execution of many independent jobs on large compute clusters. It combines two techniques to achieve this goal: (1) multi-level scheduling with separate treatments of resource provisioning and the dispatch of user tasks to those resources [23, 24], and a streamlined task dispatcher used to achieve order-of-magnitude higher task dispatch rates than conventional schedulers [3] ; and (2) data caching and the use of a data-aware scheduler to leverage the co-located computational and storage resources to minimize the use of shared storage infrastructure [25, 26, 27]. Note that we make full use of the first technique on both the BG/P and SiCortex systems. We will investigate harnessing the second technique (data diffusion) to ensure loosely coupled applications have the best opportunity to scale to the full BG/P scale of 160K processors.

We believe the synergy found between Swift and Falkon offers a common generic infrastructures and platforms in the science domain for workflow administration, scheduling, execution, monitoring, provenance tracking etc. The science community is demanding both specialized, domain-specific languages to improve productivity and efficiency in writing concurrent programs and coordination tools, and generic platforms and infrastructures for the execution and management of large scale scientific applications, where scalability and performance are major concern. High performance computing support has become an indispensable tool to address the large storage and computing problems emerging in every discipline of 21st century e-science.

Both Swift [15, 28] and Falkon [3] have been used in a variety of environments from clusters (i.e. TeraPort [29]), to multi-site Grids (i.e. Open Science Grid [30], TeraGrid [31]), to specialized large machines (SiCortex [2]), to supercomputers (i.e. IBM BlueGene/P [1]). Large scale applications from many domains (i.e. astronomy [32, 3], medicine [34, 3, 33], chemistry [28], molecular dynamics [36], and economics [38, 37]) have been run at scales of tens of thousands of jobs on thousands of processors, with an order of magnitude larger scale on the horizon.

## 3.2 Implementation Details

Significant engineering efforts were invested to get Falkon and Swift to work on systems such as the SiCortex and the BG/P. This section discusses extensions we made to both systems, and the problems and bottlenecks they addressed.

### *3.2.1 Static Resource Provisioning*

Falkon as presented in our previous work [3] has support for dynamic resource provisioning, which allows the resource pool to grow and shrink based on load (i.e. wait queue length at the Falkon service). Dynamic resource provisioning in its current implementation depends on GRAM4 [39] to allocate resources in Grids and clusters. Neither the SiCortex nor the BG/P support GRAM4; the SiCortex uses the SLURM LRM [2] while the BG/P supports the Cobalt LRM [17]. As a first step, we implemented static resource provisioning on both of these systems through the respective LRM. With static resource provisioning the application requests a number of processors for a fixed duration; resources are allocated based on the requirements of the application. In future work, we will port our GRAM4 based dynamic resource provisioning to support SLURM and Cobalt and/or pursue GRAM support on these systems.

### *3.2.2 Alternative Implementations*

Performance depends critically on the behavior of our task dispatch mechanisms; the number of messages needed to interact between the various components of the system; and the hardware, programming language, and compiler used. We implemented Falkon in Java and use the Sun JDK to compile and run Falkon. We use the GT4 Java WS-Core to handle Web Services communications. [35] Running with Java-only components works well on typical Linux clusters and Grids, but the lack of Java on the BG/L, BG/P, and SiCortex execution environments prompted us to replace two main pieces from Falkon.

The first change was the rewrite of the Falkon executor code in C. This allowed us to compile and run on the target systems. Once we had the executor implemented in C, we adapted it to interact with the Falkon service. In order to keep the protocol as lightweight as possible, we used a simple TCP-based protocol to replace the existing WS-based protocol. The process of writing the C executor was more complicated than the Java variation, and required more debugging due to exception handling, memory allocation, array handling, etc. Table 1 has a summary of the differences between the two implementations.

**Table 1: Feature comparison between the Java Executor implementation and the new C implementation**

| Description | Java | C |
|---|---|---|
| **Robustness** | high | Medium |
| **Security** | GSITransport,<br>GSIConversation,<br>GSIMessageLevel | none<br>could support SSL |
| **Communication Protocol** | WS-based | TCP-based |
| **Error Recovery** | yes | yes |
| **Lifetime Management** | yes | no |
| **Concurrent Tasks** | yes | no |
| **Push/Pull Model** | PUSH<br>notification based | PULL |
| **Firewall** | no | yes |
| **NAT / Private Networks** | no in general<br>yes in certain cases | yes |
| **Persistent Sockets** | no - GT4.0<br>yes - GT4.2 | yes |
| **Performance** | Medium~High<br>600~3700 tasks/s | High<br>1700~3200 tasks/s |
| **Scalability** | High ~ 54K CPUs | Medium ~ 10K CPUs |
| **Portability** | medium | high (needs recompile) |
| **Data Caching** | yes | no |

It was not sufficient to change the worker implementation, as the service required corresponding revisions. In addition to the existing support for WS-based protocol, we implemented a new component called “TCPCore” to handle the TCP-based communication protocol. TCPCore is a component to manage a pool of threads that lives in the same JVM as the Falkon service, and uses in-memory notifications and shared objects to communicate with the Falkon service. In order to make the protocol as efficient as possible, we implemented persistent TCP sockets (which are stored in a hash table based on executor ID or task ID, depending on what state the task is in). Figure 3 is the TCPCore overview of the interaction between TCPCore and the Falkon service, and between TCPCore and the executors.

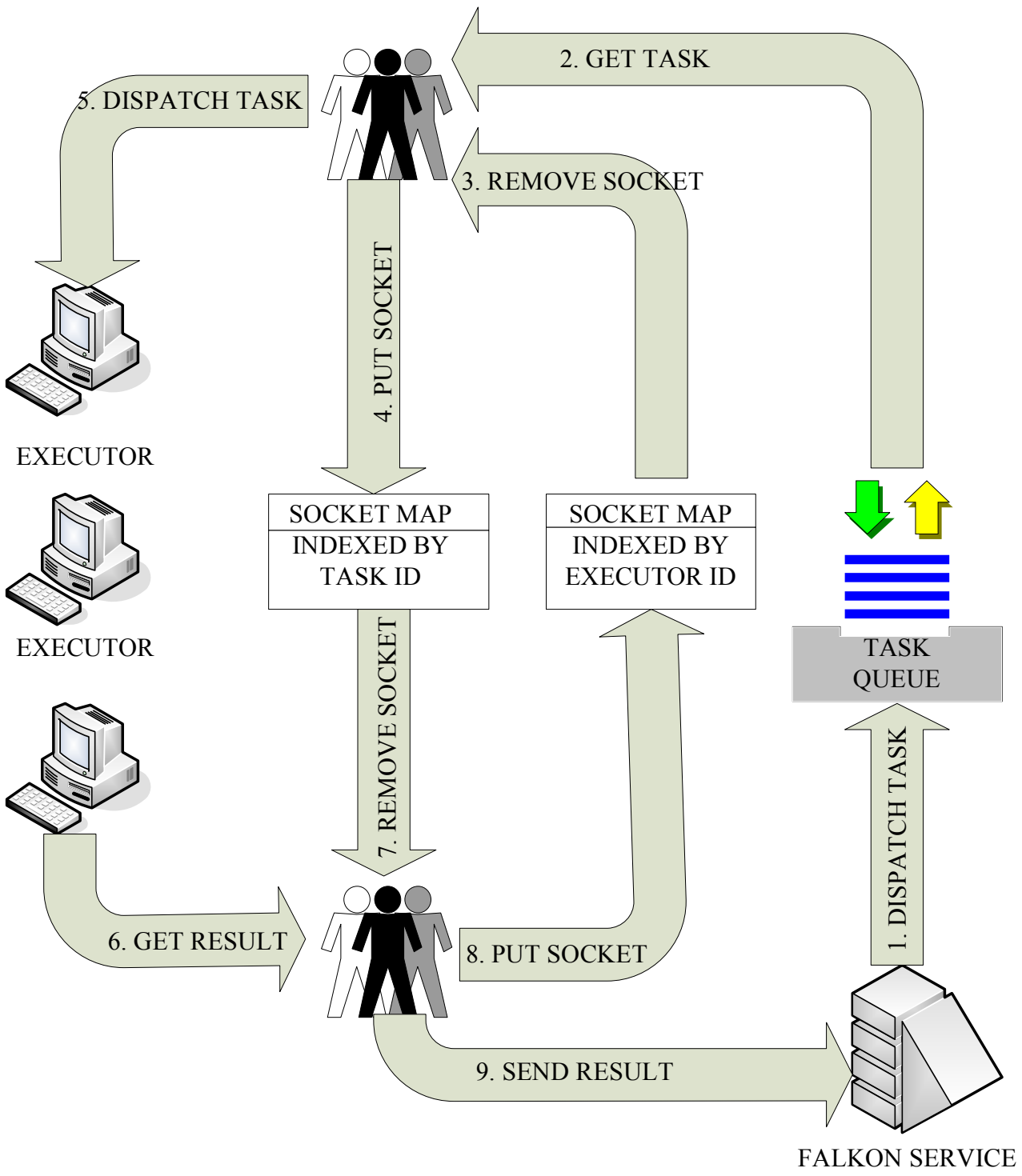

**Figure 3: The TCPCore overview, replacing the WS-Core component from the GT4**

## 3.3 Reliability Issues at Large Scale

We discuss reliability only briefly here, to explain how our approach addresses this critical requirement. The BG/L has a mean-time-to-failure (MTBF) of 10 days [4], which means that MPI parallel jobs that span more than 10 days are almost guaranteed to fail, as a single node failing would cause the entire allocation and application to fail as well. As the BG/P will scale to several orders of magnitude larger than the BG/L, we expect its MTBF to continue to pose challenges for long-running applications. When running loosely coupled applications via Swift and Falkon, the failure of a single CPU or node only affects the individual tasks that were being executed at the time of the failure.

Falkon has mechanisms to identify specific errors, and act upon them with specific actions. Most errors are generally passed back up to the application (Swift) to deal with them, but other (known) errors can be handled by Falkon. For example we have the "Stale NFS handle" error that Falkon will retry on. This error is a fail-fast error which can cause many failures to happen in a short period of time, however Falkon has the mechanisms in place to suspend the offending node if it fails too many jobs. Furthermore, Falkon retries any jobs that failed due to communication errors between the service and the workers, essentially any errors not caused due to the application or the shared file system.

Swift also has persistent state that allows it to restart a parallel application script from the point of failure, re-executing only uncompleted tasks. There is no need for explicit check-pointing as is the case with MPI applications; check-pointing occurs inherently with every task that completes and is communicated back to Swift. Compute node failures are all treated independently, as each failure only affects the particular task that it was executing at the time of failure.

# 4. Micro-Benchmarks Performance

We developed a set of micro-benchmarks to identify performance characteristics and potential bottlenecks on systems with massive numbers of cores. We describe these machines in detail and measure both the task dispatch rates we can achieve for synthetic benchmarks and the costs for various file system operations (read, read+write, invoking scripts, mkdir, etc) on the shared file systems that we use when running large-scale applications (GPFS and NFS).

## 4.1 Testbeds Description

The latest IBM BlueGene/P Supercomputer [1] has quad core processors with a total of 160K-cores, and has support for a lightweight Linux kernel (ZeptOS [19]) on the compute nodes, making it significantly more accessible to new applications. A reference BG/P with 16 PSETs (1024 nodes, 4096 processors) has been available to us for testing, and the full 640 PSET BG/P will be online at Argonne National Laboratory (ANL) later this year [20]. The BG/P architecture overview is depicted in Figure 4.

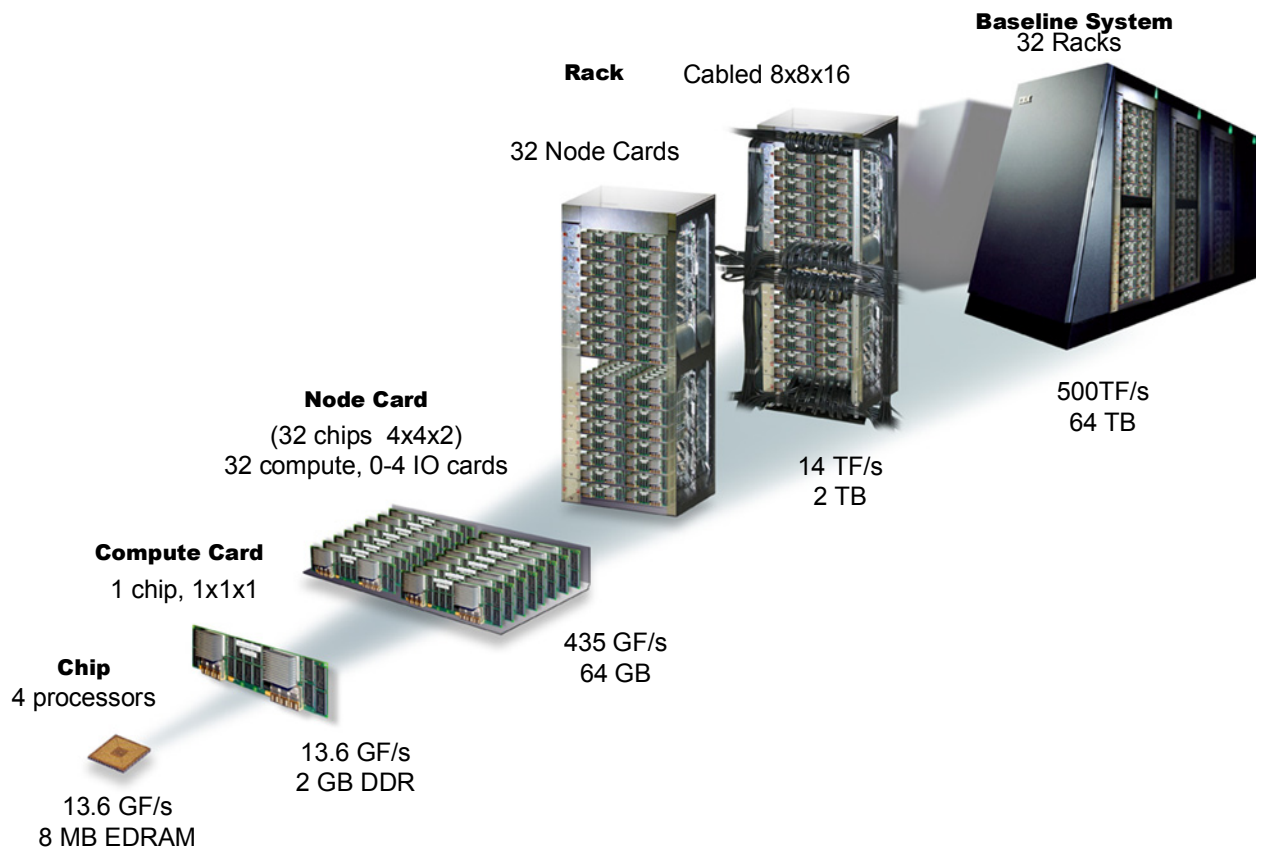

**Figure 4: BG/P Architecture Overview**

The full BG/P at ANL will be 111 TFlops with 160K PPC450 processors running at 850MHz, with a total of 80 TB of main memory. The system architecture is designed to scale to 3 PFlops for a total of 4.3 million processors. The BG/P has both GPFS and PVFS file systems available; in the final production system, the GPFS will be able to sustain 80Gb/s I/O rates. All experiment involving the BG/P were performed on the reference pre-production implementation that had 4096 processors and using the GPFS shared file system.

ANL also acquired a new 6 TFlop machine named the SiCortex [2]; it has 6-core processors for a total of 5832-cores each running at 500 MHz, has a total of 4TB of memory, and runs a standard Linux environment with kernel 2.6. The system is connected to a NFS shared file system which is only served by one server, and can sustain only about 320 Mb/s read performance. A PVFS shared file system is also planned that will increase the read performance to 12,000 Mb/s, but that was not available to us during out testing. All experiment on the SiCortex were performed using the NFS shared file system, the only available shared file system at the time of the experiments.

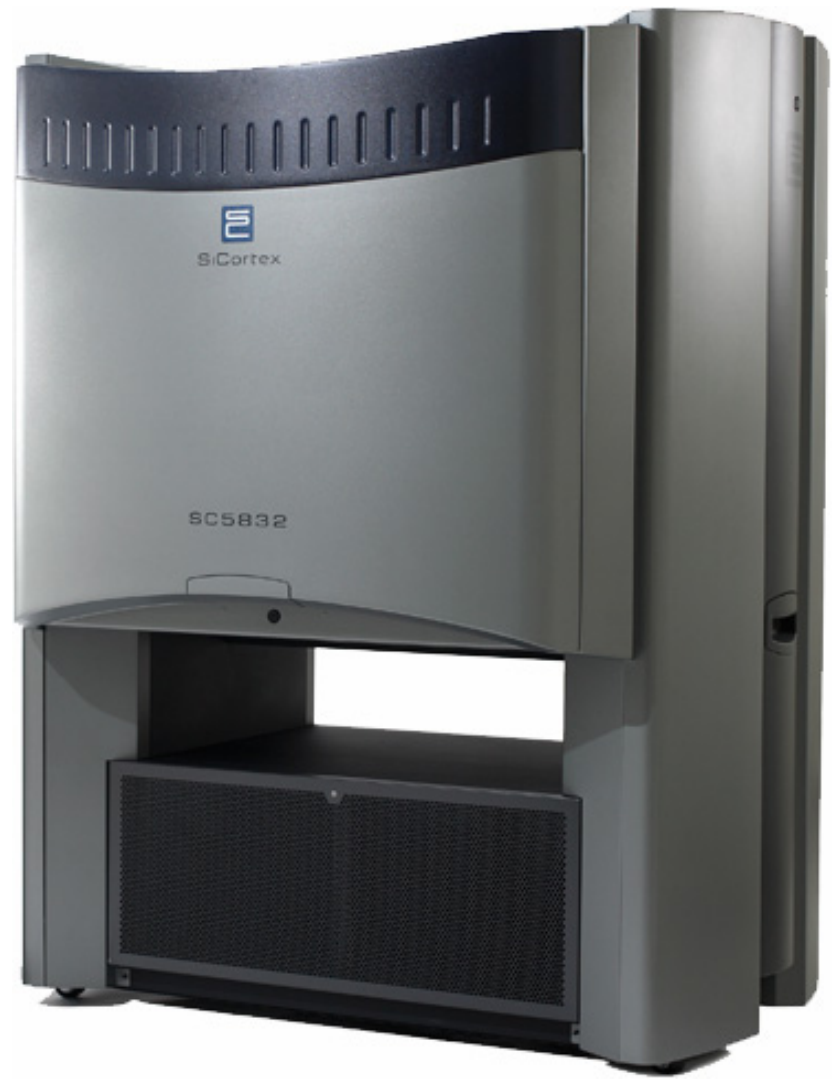

**Figure 5: SiCortex Model 5832**

In some experiments, we also used the ANL/UC Linux cluster (a 128 node cluster from the TeraGrid), which consisted of dual Xeon 2.4 GHz CPUs or dual Itanium 1.3GHz CPUs with 4GB of memory and 1Gb/s network connectivity. We also used two other systems for some of the measurements involving the SiCortex and the ANL/UC Linux cluster. One machine (VIPER.CI) was a dual Xeon 3GHz with HT (4 hardware threads), 2GB of RAM, Linux kernel 2.6, and 100 Mb/s network connectivity. The other system (GTO.CI) was a dual Xeon 2.33 GHz with quad cores each (8-cores), 2GB of RAM, Linux kernel 2.6, and 100 Mb/s network connectivity. Both machines had a network latency of less than 2 ms to and from both the SiCortex compute nodes and the ANL/UC Linux cluster.

The various systems we used in the experiments conducted in this paper are outlined in Table 2.

**Table 2: Summary of testbeds used in section 5 and section 6**

| Name | Nodes CPUs | CPU Type Speed | RAM | File System Peak | Operating System |
|---|---|---|---|---|---|
| BG/P | 1024 4096 | PPC450 0.85GHz | 2TB | GPFS 775Mb/s | Linux (ZeptOS) |
| BG/P.Login | 8 32 | PPC 2.5GHz | 32GB | GPFS 775Mb/s | Linux Kernel 2.6 |
| SiCortex | 972 5832 | MIPS64 0.5GHz | 3.5TB | NFS 320Mb/s | Linux Kernel 2.6 |
| ANL/UC | 98 196 | Xeon 2.4GHz | 0.4TB | GPFS 3.4Gb/s | Linux Kernel 2.4 |
| | 62 124 | Itanium 1.3GHz | 0.25TB | GPFS 3.4Gb/s | Linux Kernel 2.4 |
| VIPER.CI | 1 2 | Xeon 3GHz | 2GB | Local 800Mb/s | Linux Kernel 2.6 |
| GTO.CI | 1 8 | Xeon 2.3GHz | 2GB | Local 800Mb/s | Linux Kernel 2.6 |

## 4.2 Falkon Task Dispatch Performance

One key component to achieving high utilization of large scale systems is the ability to get high dispatch and execute rates. In previous work [3] we measured that Falkon with the Java Executor and WS-based communication protocol achieves 487 tasks/sec in a Linux cluster (ANL/UC) with 256 CPUs, where each task was a “sleep 0” task with no I/O; the machine we used in our previous study was VIPER.CI. We repeated the peak throughput experiment on a variety of systems (ANL/UC Linux cluster, SiCortex, and BG/P) for both versions of the executor (Java and C, WS-based and TCP-base respectively); we also used two different machines to run the service, GTO.CI and BG/P.Login; see Table 2 for a description of each machine.

Figure 6 shows the results we obtained for the peak throughput as measured while submitting, executing, and getting the results from 100K tasks on the various systems. We see that the ANL/UC Linux cluster is up to 604 tasks/sec from 487 tasks/sec (using the Java executor and the WS-based protocol); we attribute the gain in performance solely due to the faster machine GTO.CI (8-cores at 2.33GHz vs. 2 CPUs with HT at 3GHz each). The test was performed on 200 CPUs, the most CPUs that were available at the time of the experiment. The same testbed but using the C executor and TCP-based protocol yielded 2534 tasks/sec, a significant improvement in peak throughput. We attribute this to the lesser overhead of the TCP-based protocol (as opposed to the WS-based protocol), and the fact that the C executor is much simpler in logic and features than the Java executor. The same peak throughput on the SiCortex with 5760 CPUs is even higher, 3186 tasks/sec; note that the SiCortex does not support Java. Finally, the BG/P peak throughput was only 1758 tasks/sec for the C executor; similar to the SiCortex, Java is not supported on the BG/P compute nodes. We attribute the lower throughput of the BG/P as compared to the SiCortex to the machine that was used to run the Falkon service. On the BG/P, we used BG/P.Login (a 4-core PPC at 2.5GHz) while on the SiCortex we used GTO.CI (a 8-core Xeon at 2.33GHz). These differences in test harness were unavoidable due to firewall constraints.

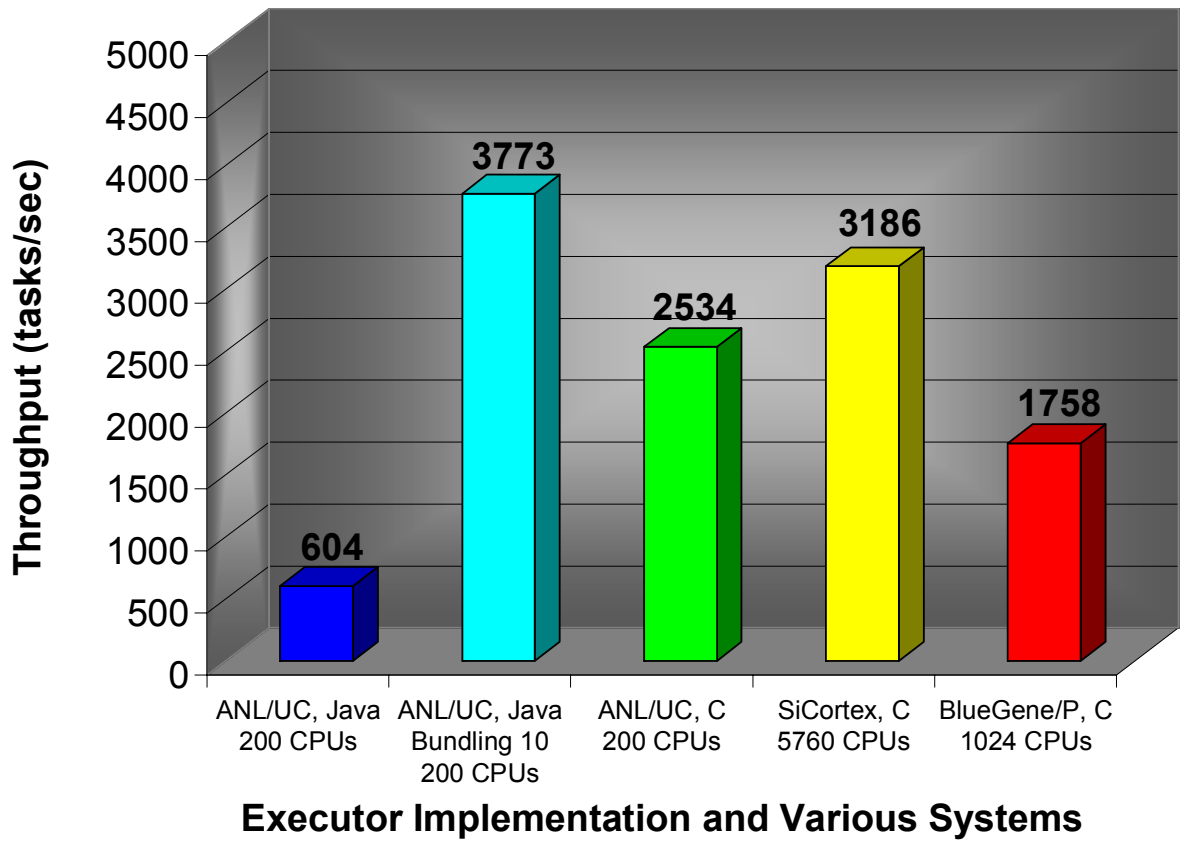

**Figure 6: Task dispatch and execution throughput for trivial tasks with no I/O (sleep 0)**

Note that there is also an entry for the ANL/UC Linux cluster, with the Java executor and bundling attribute of 10. The bundling refers to the dispatcher bundling 10 tasks in each communication message that is sent to a worker; the worker then unbundles the 10 tasks, puts them in a local queue, and executes one task per CPU (2 in our case) at a time. This has the added benefit of amortizing the communication overhead over multiple tasks, which drastically improves throughput from 604 to 3773 tasks/sec (higher than all the C executors and TCP-based protocol). Bundling can be useful when one knows a-priori the task granularity, and expects the dispatch throughput to be a bottleneck. The bundling feature has not been implemented in the C executor, which means that the C executors were receiving each task separately per executor.

In trying to understand the various costs leading to the throughputs achieved in Figure 6, Figure 7 profiles the service code, and breaks down the CPU time by code block. This test was done on the VIPER.CI and the ANL/UC Linux cluster with 200 CPUs, with throughputs reaching 487 tasks/sec and 1021 tasks/sec for the Java and C implementations respectively. A significant portion of the CPU time is spent in communication (WS and/or TCP). With bundling (not shown in Figure 7), the communication costs are reduced to 1.2 ms (down from 4.2 ms), as well as other costs. Our conclusion is that the peak throughput for small tasks can be increased by both adding faster processors, more processor cores to the service host, and reducing the communication costs by lighter weight protocols or by bundling where possible.

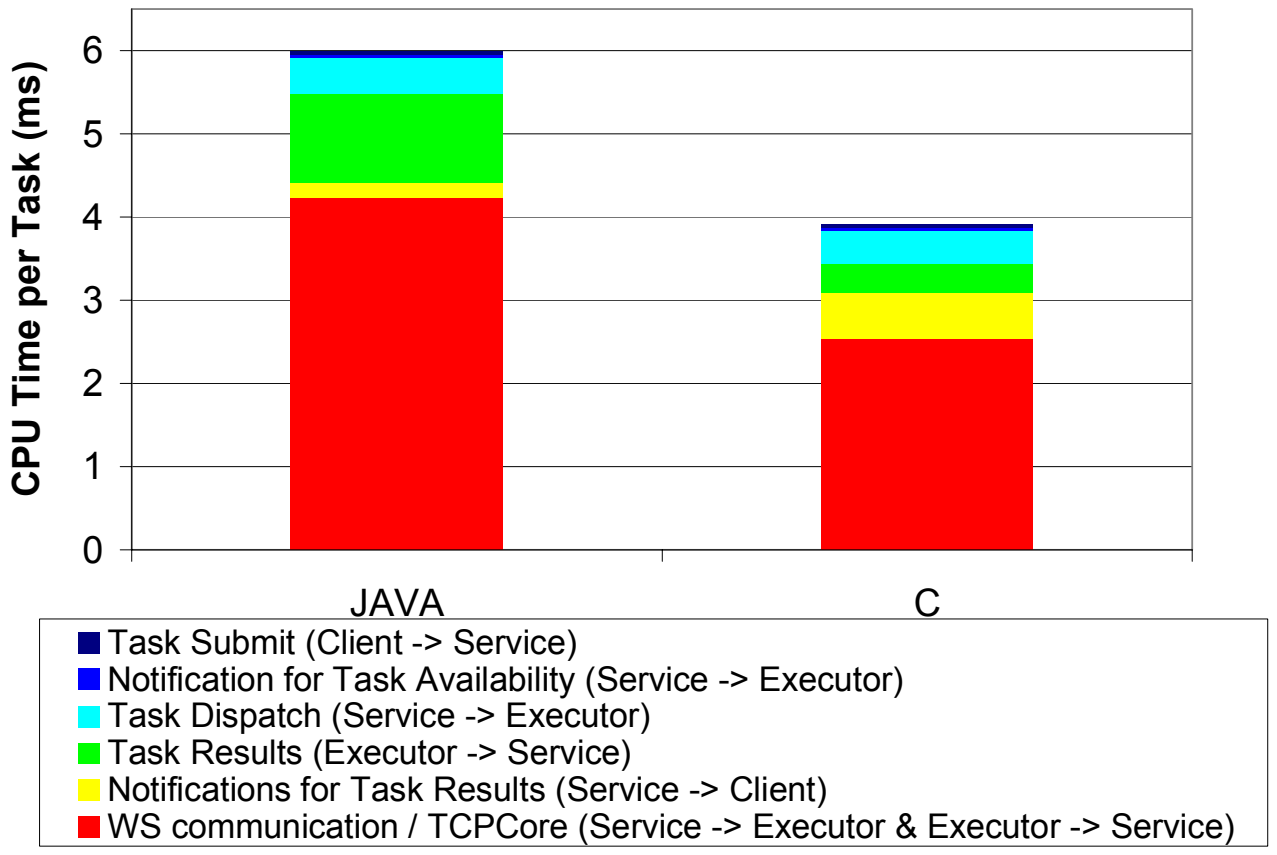

**Figure 7: Falkon profiling comparing the Java and C implementation on VIPER.CI (dual Xeon 3GHz w/ HT)**

Peak throughput performance only gives us a rough idea of the kind of utilization and efficiency we can expect; therefore, to better understand the efficiency of executing different workloads, measured the efficiency of executing varying task lengths. We measured on the ANL/UC Linux cluster with 200 CPUs, the SiCortex with 5760 CPUs, and the BG/P with 2048 CPUs. We varied the task lengths from 0.1 seconds to 256 seconds (using sleep tasks with no I/O), and ran workloads ranging from 1K tasks to 100K tasks (depending on the task lengths).

Figure 8 shows the efficiency we were able to achieve. Note that on a relatively small cluster (200 CPUs), we can achieve 95%+ efficiency with 1 second tasks. Even with 0.1 second tasks, using the C executor, we can achieve 70% efficiency on 200 CPUs. Efficiency can reach 99%+ with 16 second tasks. With larger systems, with more CPUs to keep busy, it takes longer tasks to achieve a given efficiency level. With 2048 CPUs (BG/P), we need 4 second tasks to reach 94% efficiency, while with 5760 CPUs (SiCortex), we need 8 second tasks to reach the same efficiency. With 64 second tasks, the BG/P achieves 99.1% efficiency while the SiCortex achieves 98.5%.

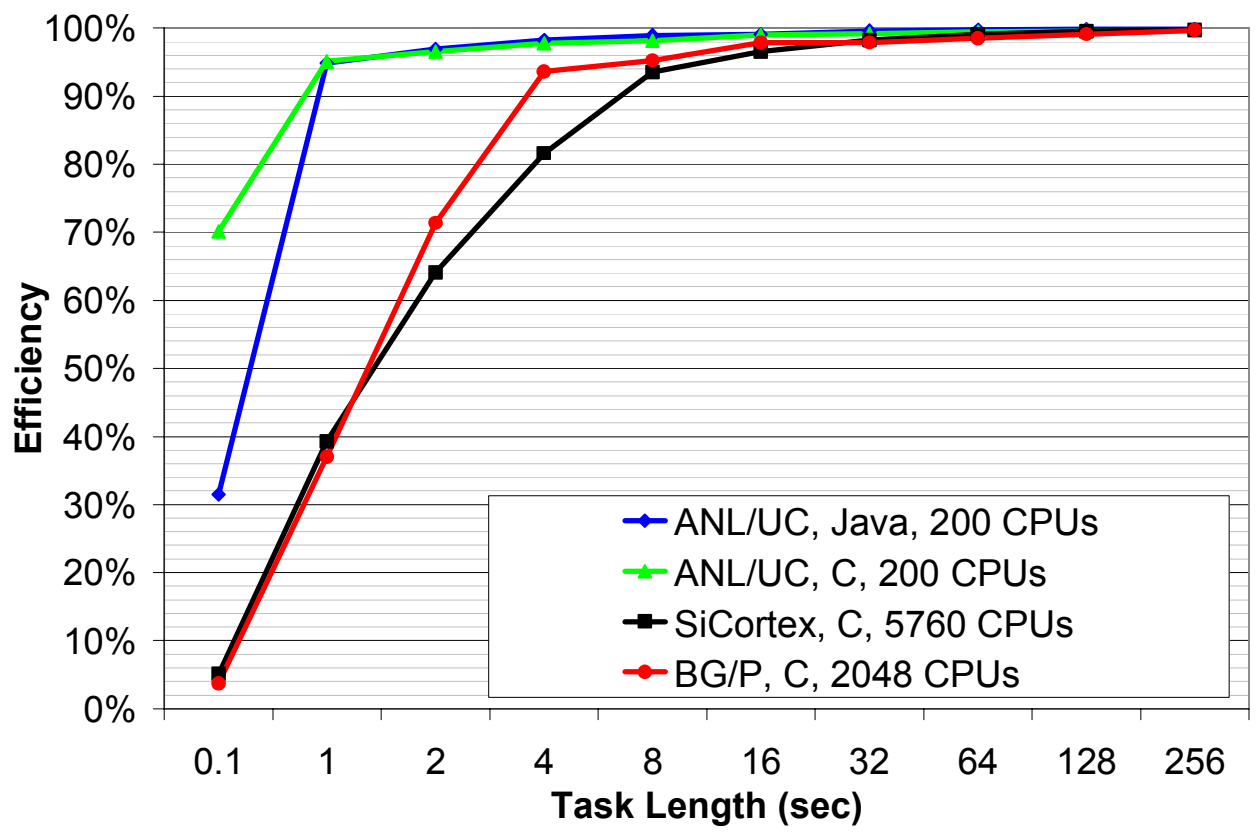

**Figure 8: Efficiency graph of various systems (BG/P, SiCortex, and Linux cluster) for both the Java and C worker implementation for various task lengths (0.1 to 256 seconds)**

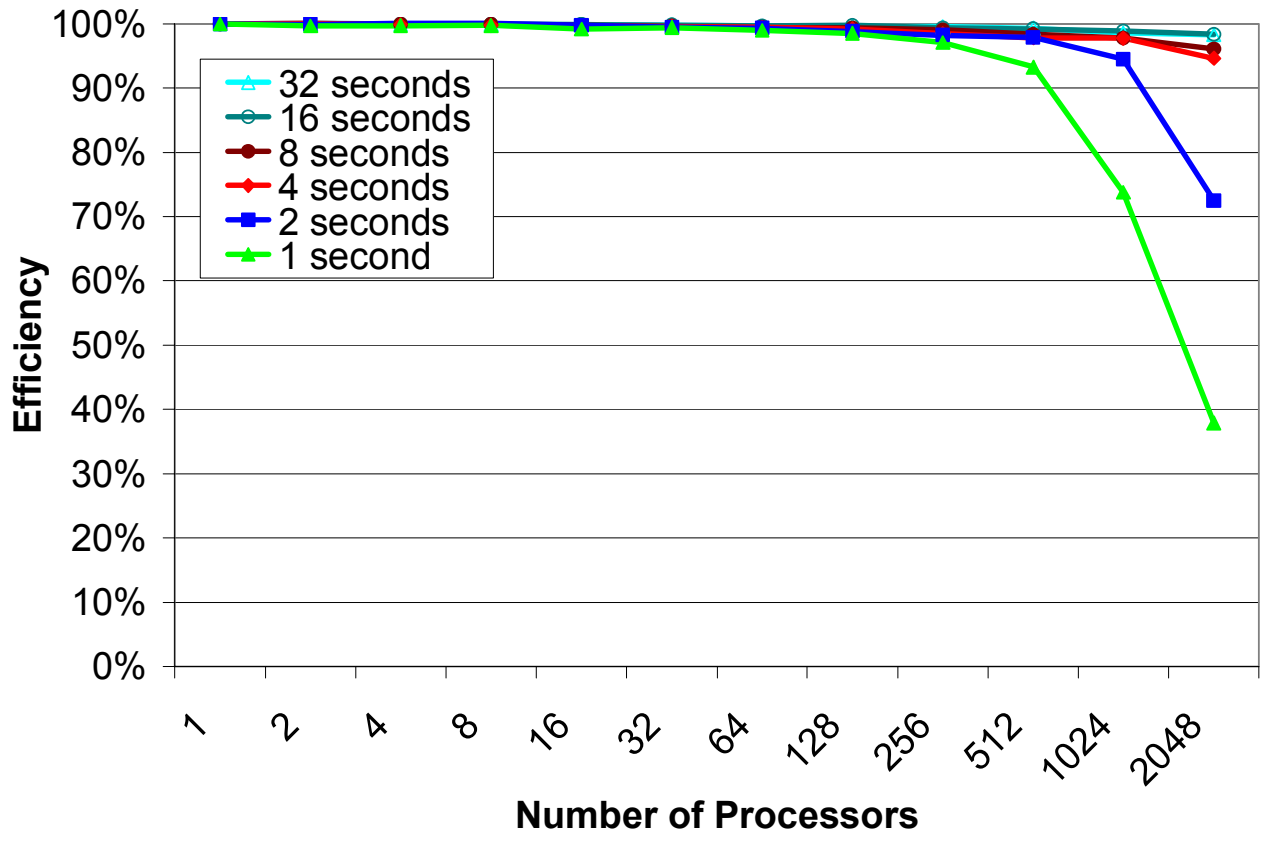

**Figure 9: Efficiency graph for the BG/P for 1 to 2048 processors and task lengths ranging from 1 to 32 seconds**

Figure 9 investigates more closely the effects of efficiency as the number of processors increases from 1 to 2048. With 4 second tasks, we can get high efficiency with any number of processors; with 1 and 2 second tasks, we achieve high efficiency with a smaller number of processors: 512 and 1024 respectively.

The previous several experiments all investigated the throughput and efficiency of executing tasks which had a small and compact description. For example, the task "/bin/sleep 0" requires only 12 bytes of information. The following experiment (Figure 10) investigates how the throughput is affected by increasing the task description size. For this experiment, we compose 4 different tasks, "/bin/echo 'string'", where string is replaced with a different length string to make the task description 10B, 100B, 1KB, and 10KB. We ran this experiment on the SiCortex with 1002 CPUs and the service on GTO.CI, and processed 100K tasks for each case.

We see the throughput with 10B tasks is similar to that of sleep 0 tasks on 5760 CPUs with a throughput of 3184 tasks/sec. When the task size is increased to 100B, 1KB, and 10KB, the throughput is reduced to 3011, 2001, and 662 tasks/sec respectively. To better understand the throughput reduction, we also measured the network level traffic that the service experienced during the experiments. We observed that the aggregate throughput (both received and sent on a full duplex 100Mb/s network link) increases from 2.9MB/s to 14.4MB/s as we vary the task size from 10B to 10KB.

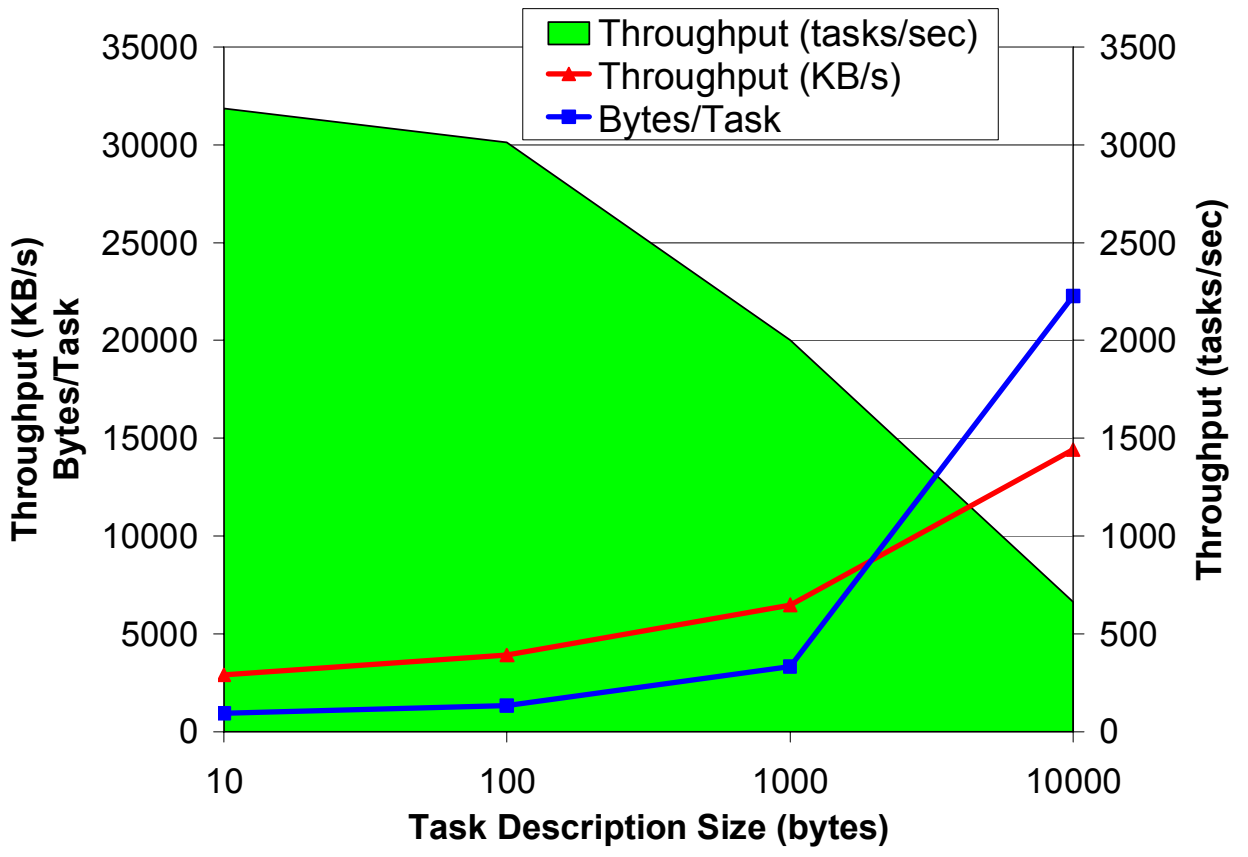

**Figure 10: Task description size on the SiCortex and 1K CPUs**

The bytes/task varies from 934 bytes to 22.3 KB for the 10B to 10KB tasks. The formula to compute the bytes per task is 2*task_size + overhead of TCP-based protocol (including TCP/IP headers) + overhead of WS-based submission protocol (including XML, SOAP, HTTP, and TCP/IP) + notifications of results from executors back to the service, and from the service to the user. We need to double the task size since the service first receives the task description from the user (or application), and then dispatches it to the remote executor. Only a brief notification with the task ID and exit code of the application is sent back. We might assume that the overhead is 934 – 2*10 = 914 bytes, but from looking at the 10KB tasks, we see that the overhead is 22.3KB – 2*10KB = 2.3KB (higher than 0.9KB). We measured the number of TCP packets to be 7.36 packets/task (10B tasks) and 28.67 packets/task (10KB tasks). The difference in TCP overhead 853 bytes (with 40 byte headers for TCP/IP, 28.67*40 - 7.36*40) explains most of the difference. We suspect that the remainder of the difference (513 bytes) is due to extra overhead in XML/SOAP/HTTP when submitting the tasks.

## 4.3 NFS/GPFS Performance

Another key component to getting high utilization and efficiency on large scale systems is to understand the shared resources well, and to make sure that the compute-to-I/O ratio is proportional in order to achieve the desired performance. This sub-section discusses the shared file system performance of the BG/P. This is an important factor, as Swift uses files for inter-process communication, and these files are transferred from one node to another by means of the shared file system. Future work will remove this bottleneck (i.e. using TCP pipes, MPI messages, or data diffusion [25, 27]), but the current implementation is based on files on shared file systems, and hence we believe it is important to investigate and measure the performance of the BG/P's GPFS shared filesystem.

We conducted several experiments with various data sizes (1B to 100MB) on a varying number of CPUs from 4 to 2048; we conducted both read-only tests and read+write tests. Figure 11 shows the aggregate throughput in terms of Mb/s. Note that it requires relatively large access sizes (1MB and larger) in order to saturate the GPFS file system (and/or the I/O nodes that handle the GPFS traffic). The peak throughput achieved for read tests was 775 Mb/s with 1MB data sizes, and 326 Mb/s read+write throughput with 10MB data sizes. At these peak numbers, 2048 CPUs are concurrently accessing the shared file system, so the peak per processor throughput is a mere 0.379 Mb/s and 0.16 Mb/s for read and read+write respectively. This implies that care must be taken to ensure that the compute to I/O ratio to and from the shared file system is balanced in such a way that it fits within the relatively low per processor throughput.

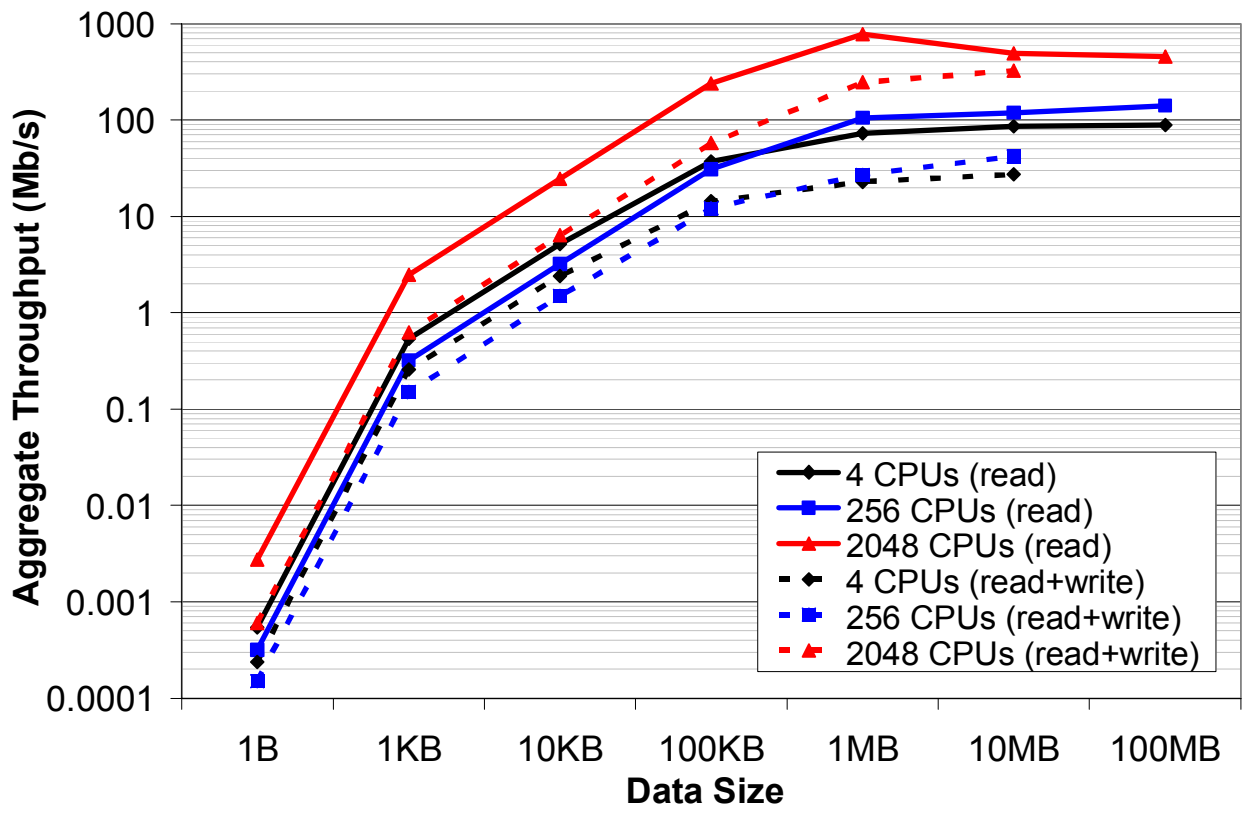

**Figure 11: Aggregate throughput for GPFS on the BG/P**

Figure 12 shows the same information as Figure 11, but shows task length necessary to achieve 90% efficiency; we show the task length required when reading from GPFS in solid lines and read+write from GPFS in dotted lines.

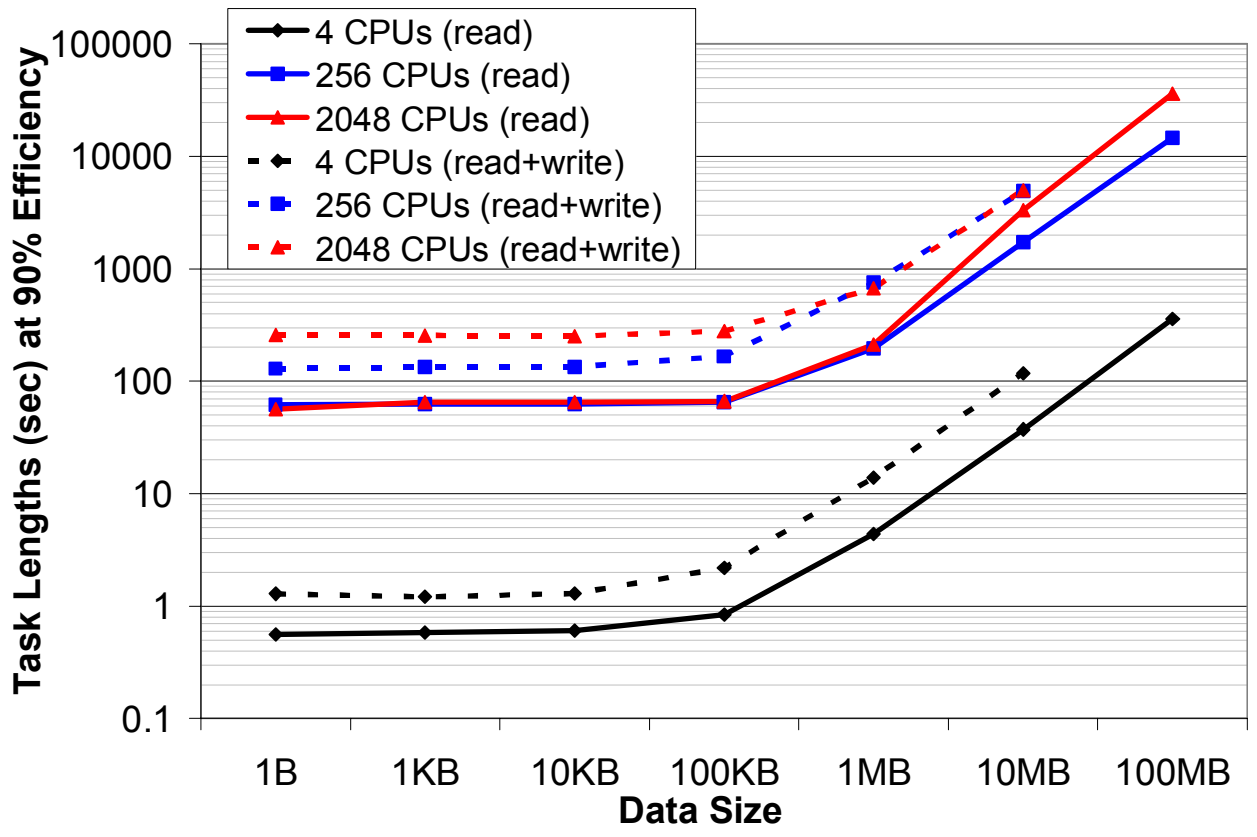

**Figure 12: Minimum task lengths (sec) with varying input data required to maintain 90% efficiency**

Looking at the measures of 1 P-SET (blue) and 8 P-SETs (red), we see that no matter how small the input/output data is (1B ~ 100KB), we need to have at least 60+ second tasks to achieve 90% efficiency. If we do both reads and writes, we need at least 129 sec tasks and 260 sec tasks for the 1 byte case for read and read+write respectively. This paints a bleak picture of the BG/P's performance when we need to access GPFS. It is essential that these ratios (task length vs. data size) be considered when implementing an application on the BG/P which needs to access the data from the shared file system (using the loosely coupled model under consideration).

Figure 13 shows another aspect of the GPFS performance on the BG/P for 3 different scales, 4, 256, and 2048 processors. It investigates 2 different benchmarks, the speed at which scripts can be invoked from GPFS, and the speed to create and remove directories on GPFS. We show both aggregate throughput and time (ms) per operation per processor.

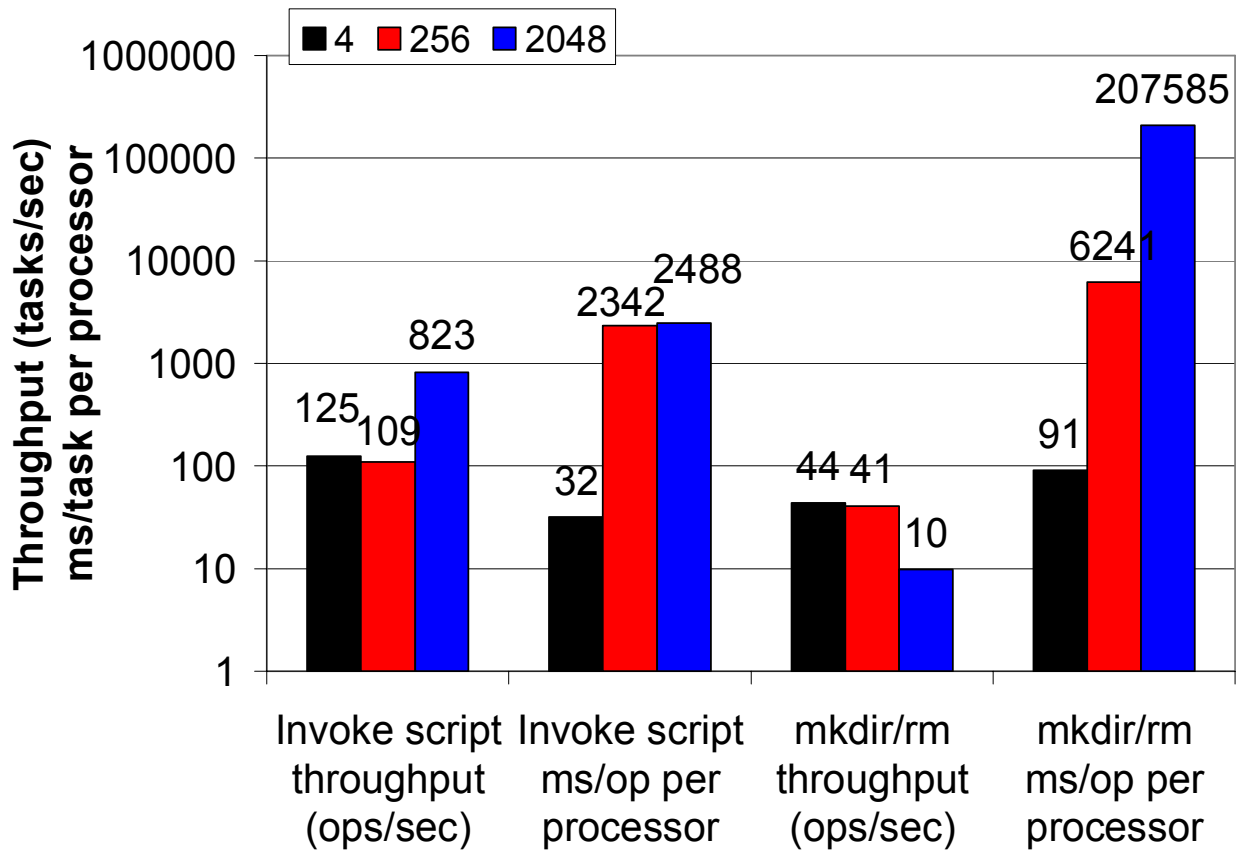

**Figure 13: invoking simple script and mkdir/rm**

Looking at the script invocation first (left two columns), we see that we can only invoke scripts at 109 tasks/sec with 256 processors; it is interesting to note that if this script was on ramdisk, we can achieve over 1700 tasks/sec. As we add extra processors up to 2048 (also increasing I/O nodes from 1 to 8), we get almost linear increase in throughput, with 823 tasks/sec. This leads us to conclude that the I/O nodes are the main bottleneck for invoking scripts from GPFS, and not GPFS itself. Also, note the time increase per script invocation per processor, going from 4 to 256 processors increases from 32 ms to 2342 ms, a significant overhead for relatively small tasks. The second microbenchmark investigated the performance of creating and removing directories (right two columns). We see that the aggregate throughput stays relatively constant with 4 and 256 processors (within 1 PSET) at 44 and 41 tasks/sec, but drops significantly to 10 tasks/sec with 2048 processors. Note at 2048 processors, the time needed per processor to create and remove a directory on GPFS is over 207 seconds, an extremely large overhead in comparison with a ramdisk create/remove directory overhead that is in the range of milliseconds.

It is likely that these numbers will improve with time, as the BG/P moves from an early testing machine to a full-scale production system. For example, the peak advertised GPFS performance is rated at 80Gb/s, yet we only achieved 0.77Gb/s. We only used 2048 processors (of the total 160K processors that will eventually make up the ALCF BG/P), so if GPFS scales linearly, we will achieve 61.6 Gb/s. It is possible that in the production system with 160K processors, we will not require the full machine to achieve the peak shared file system throughput (as is typical in most large clusters with shared file systems).

# 5. Loosely Coupled Applications

Synthetic tests and applications offer a great way to understand the performance characteristics of a particular system, but they do not always trivially translate into predictions of how real applications with real I/O will behave. We have worked with two separate groups of scientists from different domains as a first step to show that large-scale loosely-coupled applications can run efficiently on the BG/P and the SiCortex systems. The applications are from two domains, molecular dynamics and economic modeling, and both show excellent speedup and efficiency as they scale to thousands of processors.

## 5.1 Molecular Dynamics: DOCK

Our first application is DOCK Version 5 [36], which we have run on both the BG/P and the SiCortex systems via Swift [15, 28] and Falkon [3]. DOCK addresses the problem of "docking" molecules to each other. In general, "docking" is the identification of the low-energy binding modes of a small molecule, or ligand, within the active site of a macromolecule, or receptor, whose structure is known. A compound that interacts strongly with, or binds, a receptor (such as a protein molecule) associated with a disease may inhibit its function and thus act as a beneficial drug.

Prior to running the real workload, which exhibits wide variability in its job durations, we investigated the scalability of the application under larger than normal I/O to compute ratios and by reducing the number of variables. From the ligand search space, we selected one that needed 17.3 seconds to complete. We then ran a workload with this specific molecule (replicated to many files) on a varying number of processors from 6 to 5760 on the SiCortex. The ratio of I/O to compute was about 35 times higher in this synthetic workload than the real workload whose average task execution time was 660 seconds. Figure 14 shows the results of the synthetic workload on the SiCortex system.

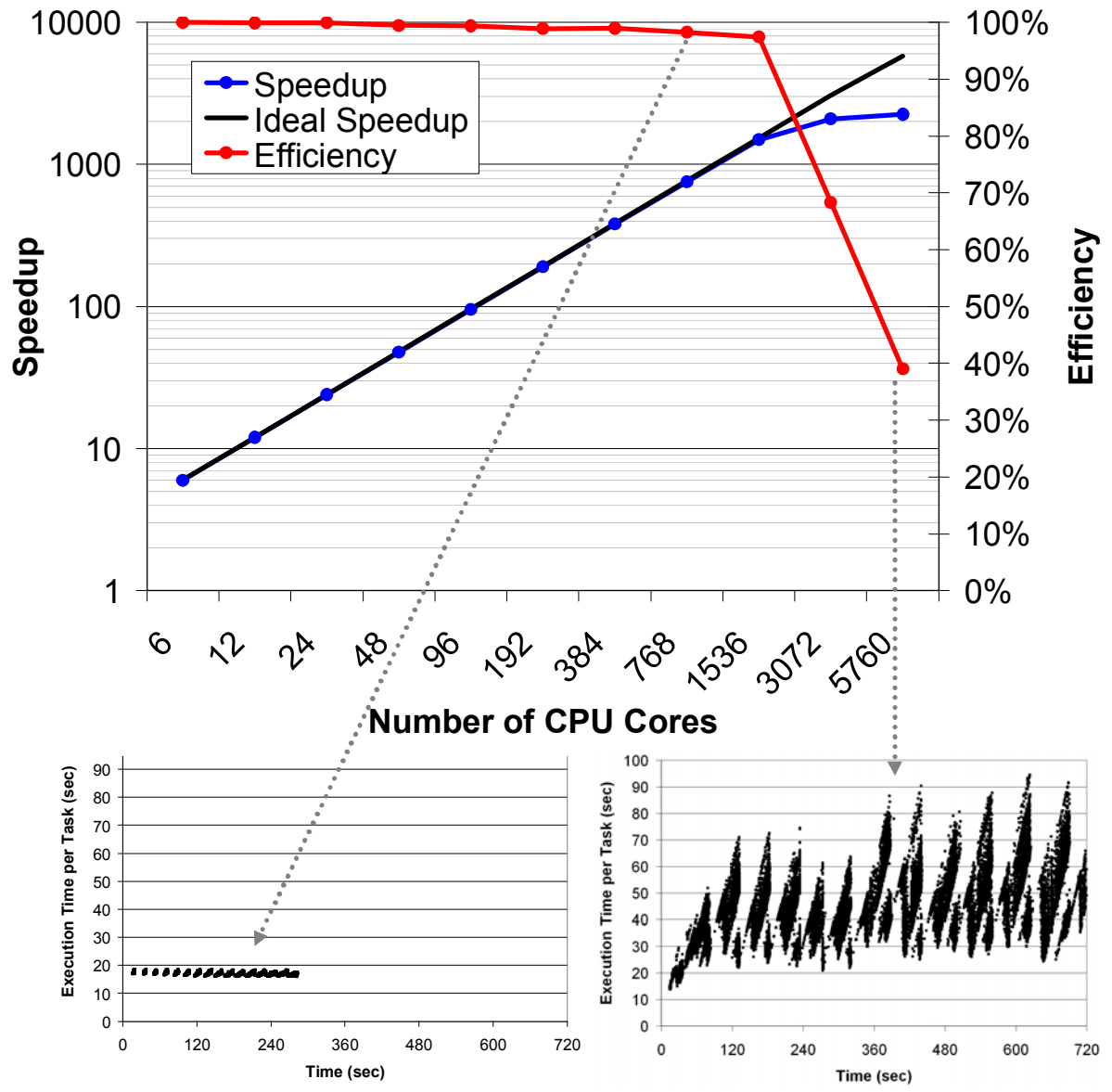

**Figure 14: Synthetic workload with deterministic job execution times (17.3 seconds) while varying the number of processors from 6 to 5760 on the SiCortex**

Up to 1536 processors, the application had excellent scalability with 98% efficiency, but due to shared file system contention in reading the input data and writing the output data, the efficiency dropped to below 70% for 3072 processors and below 40% for 5760 processors. We concluded that shared file system contention caused the loss in efficiency, due to the average execution time per job and the standard deviation as we increased the number of processors. Notice in the lower left corner of Figure 14 how stable the execution times are when running on 768 processors, 17.3 seconds average and 0.336 seconds standard deviation. However, the lower right corner shows the performance on 5760 processors to be an average of 42.9 seconds, and a standard deviation of 12.6 seconds. Note that we ran another synthetic workload that had no I/O (sleep 18) at the full 5760 processor machine scale, which showed an average of 18.1 second execution time (0.1 second standard deviation), which rules out the dispatch/execute mechanism. The likely contention was due to the application's I/O patterns to the shared file system.

The real workload of the DOCK application involves a wide range of job execution times, ranging from 5.8 seconds to 4178 seconds, with a standard deviation of 478.8 seconds. This workload (Figure 15 and Figure 16) has a 35X smaller I/O to compute ratio than the synthetic workload presented in Figure 14. Expecting that the application would scale to 5760 processors, we ran a 92K job workload on 5760 processors. In 3.5 hours, we consumed 1.94 CPU years, and had 0 failures throughout the execution of the workload. We also ran the same workload on 102 processors to compute speedup and efficiency, which gave the 5760 processor experiment a speedup of 5650X (ideal being 5760) and an efficiency of 98.2%. Each horizontal green line represents a job computation, and each black tick mark represents the beginning and end of the computation. Note that a large part of the efficiency was lost towards the end of the experiment as the wide range of job execution times yielded the slow ramp-down of the experiment and leaving a growing number of processors idle.

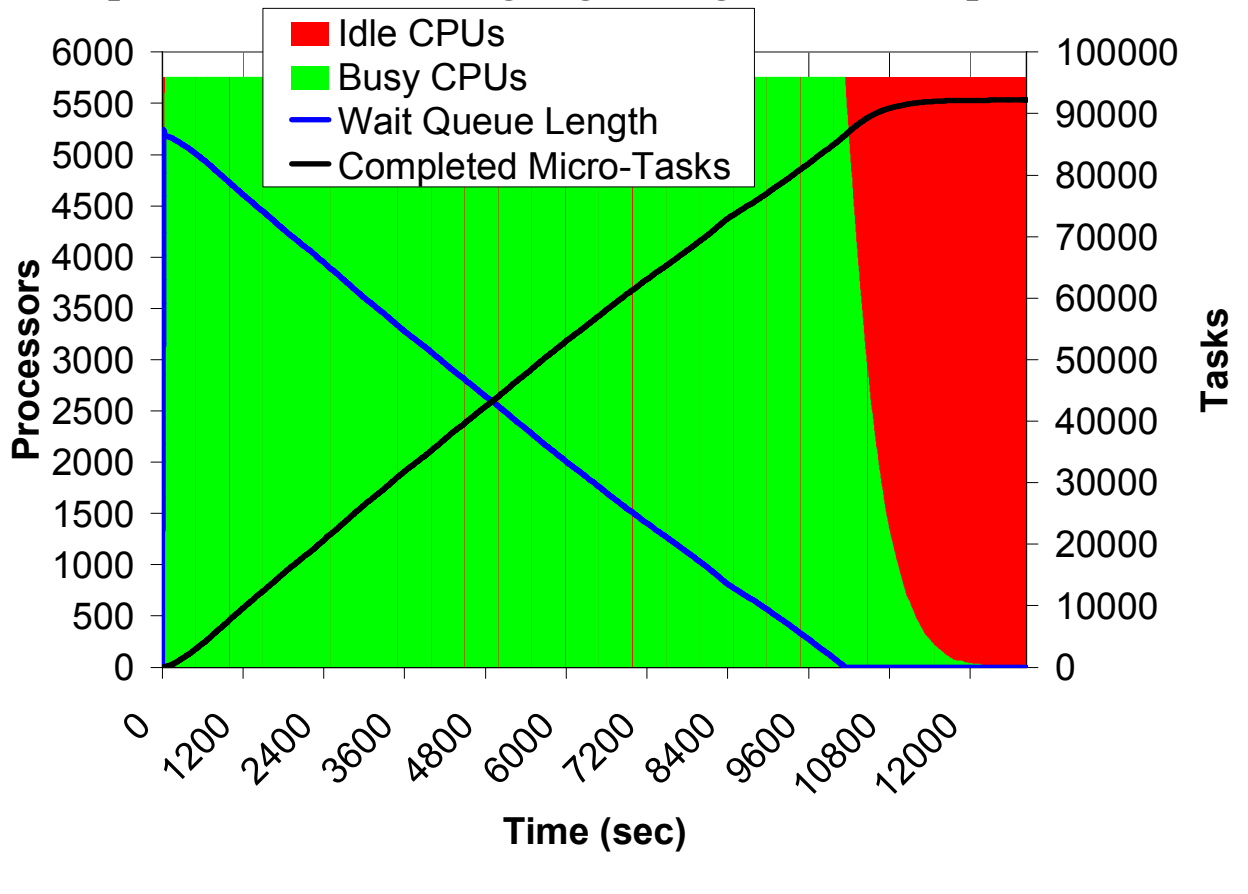

**Figure 15: DOCK application (summary view) on the SiCortex; 92K jobs using 5760 processor cores**

Despite the loosely coupled nature of this application, our preliminary results show that the DOCK application performs and scales well on thousands of processors. The excellent scalability (98% efficiency when comparing the 5760 processor run with the same workload executed on 102 processors) was achieved only after careful consideration was taken to avoid the shared file system, which included the caching of the multi-megabyte application binaries, and the caching of 35MB of static input data that would have otherwise been read from the shared file system for each job. Note that each job still had some minimal read and write operations to the shared file system, but they were on the order of 10s of KB, with the majority of the computations being in the 100s of seconds, with an average of 660 seconds.

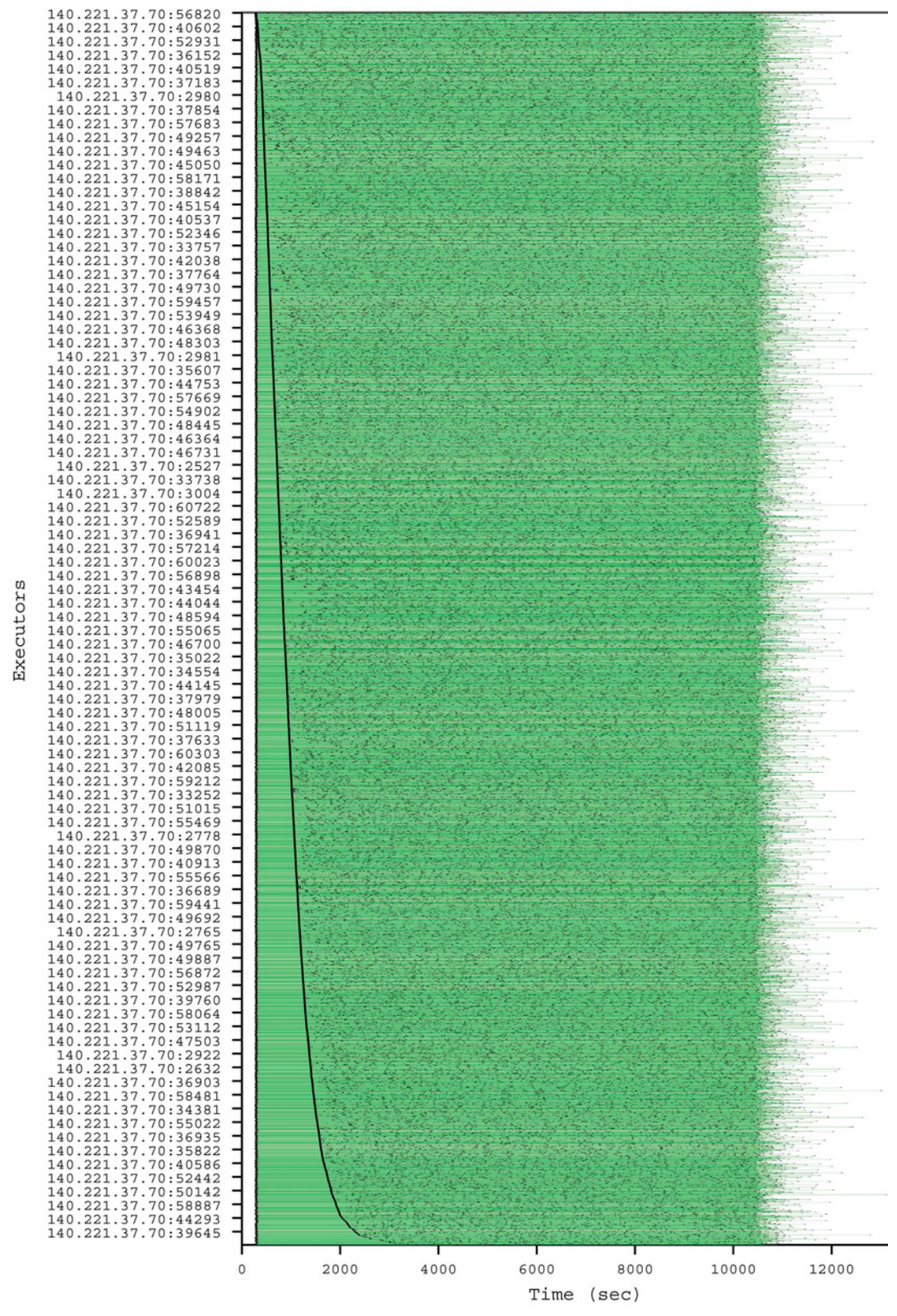

**Figure 16: DOCK application (per processor view) on the SiCortex; 92K jobs using 5760 processor cores**

To grasp the magnitude of DOCK application, the 92K jobs we performed represents only 0.0092% of the search space being considered by the scientists we are working with; simple calculations projects a search over the entire parameter space to need 20,938 CPU years, the equivalent of 4.9 years on today's 4K CPU BG/P, or 48 days on the 160K-core BG/P that will be online later this year at Argonne National Laboratory. This is a large problem, that cannot be solved in a reasonable amount of time (<1 year) without a system that has at least 10K processors or more, but our loosely-coupled approach holds great promise for making this problem tractable and manageable.

## 5.2 Economic Modeling: MARS

The second application whose performance we evaluated on our target architectures was MARS – the Macro Analysis of Refinery Systems, an economic modeling application for petroleum refining developed by D. Hanson and J. Laitner at Argonne [38]. This modeling code performs a fast but broad-based simulation of the economic and environmental parameters of petroleum refining, covering over 20 primary & secondary refinery processes. MARS analyzes the processing stages for six grades of crude oil (from low-sulfur light to high-sulfur very-heavy and synthetic crude), as well as processes for upgrading heavy oils and oil sands. It includes eight major refinery products including gasoline, diesel and jet fuel, and evaluates ranges of product shares. It models the economic and environmental impacts of the consumption of natural gas, the production and use of hydrogen, and coal-to-liquids co-production, and seeks to provide insights into how refineries can become more efficient through the capture of waste energy.

While MARS analyzes this large number of processes and variables, it does so at a coarse level without involving intensive numerics. It consists of about 16K lines of C code, and can process one iteration of a model execution in about 0.5 seconds of BG/P CPU time. Using the power of the BG/P we can perform detailed multi-variable parameter studies of the behavior of all aspects of petroleum refining covered by MARS.

As a simple test of utilizing the BG/P for refinery modeling, we performed a 2D parameter sweep to explore the sensitivity of the investment required to maintain production capacity over a 4-decade span on variations in the diesel production yields from low sulfur light crude and medium sulfur heavy crude oils. This mimics one possible segment of the many complex multivariate parameter studies that become possible with ample computing power. A single MARS model execution involves an application binary of 0.5MB, static input data of 15KB, 2 floating point input variables and a single floating point output variable. The average micro-task execution time is 0.454 seconds. To scale this efficiently, we performed task-batching of 144 model runs into a single task, yielding a workload with 1KB of input and 1KB of output data, and an average execution time of 65.4 seconds.

We executed a workload with 7 million model runs (49K tasks) on 2048 processors on the BG/P (Figure 17 and Figure 18). The experiment consumed 894 CPU hours and took 1601 seconds to complete. At the scale of 2048 processors, the per micro-task execution times were quite deterministic with an average of 0.454 seconds and a standard deviation of 0.026 seconds; this can also be seen from Figure 18 where we see all processors start and stop executing tasks at about the same time, the banding effects in the graph) . As a comparison, a 4 processor experiment of the same workload had an average of 0.449 seconds with a standard deviation of 0.003 seconds. The efficiency of the 2048 processor run in comparison to the 4 processor run was 97.3% with a speedup of 1993 (compared to the ideal speedup of 2048).

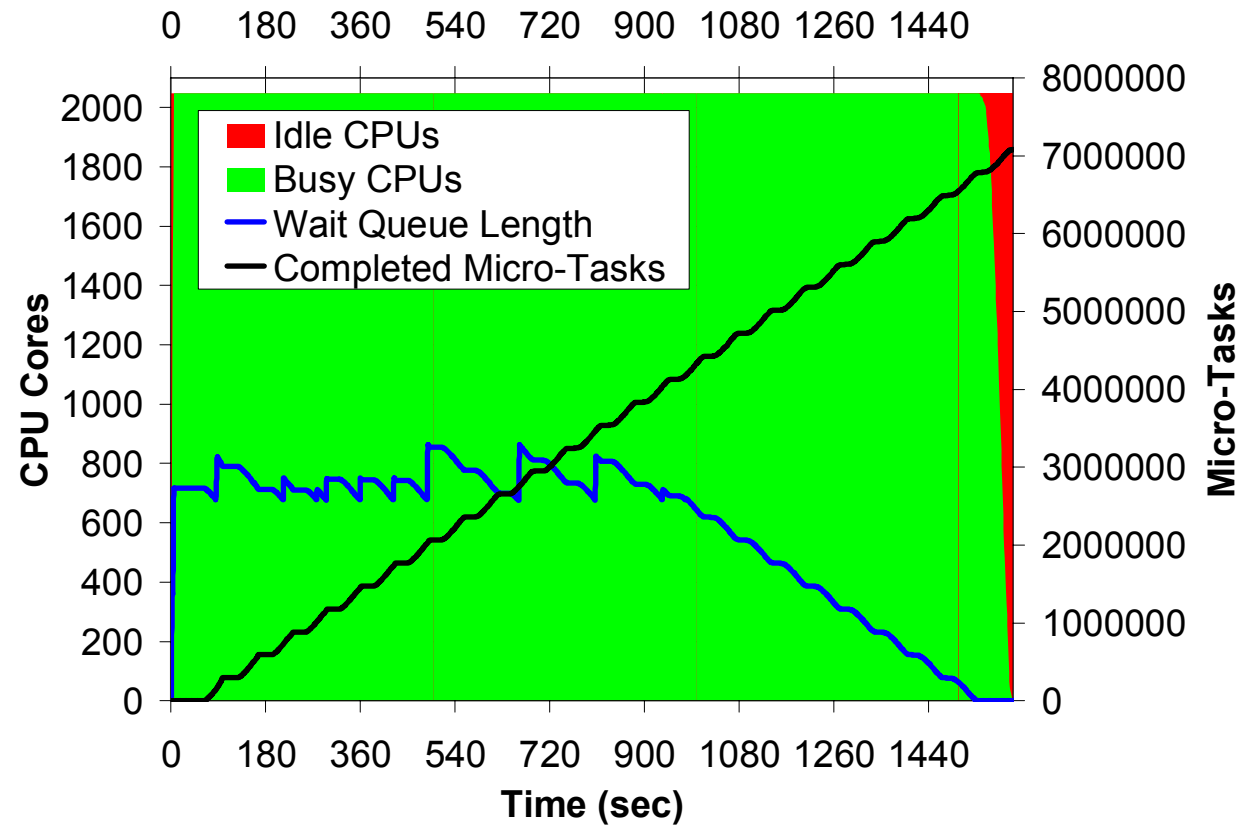

**Figure 17: MARS application (summary view) on the BG/P; 7M micro-tasks (49K tasks) using 2048 processor cores**

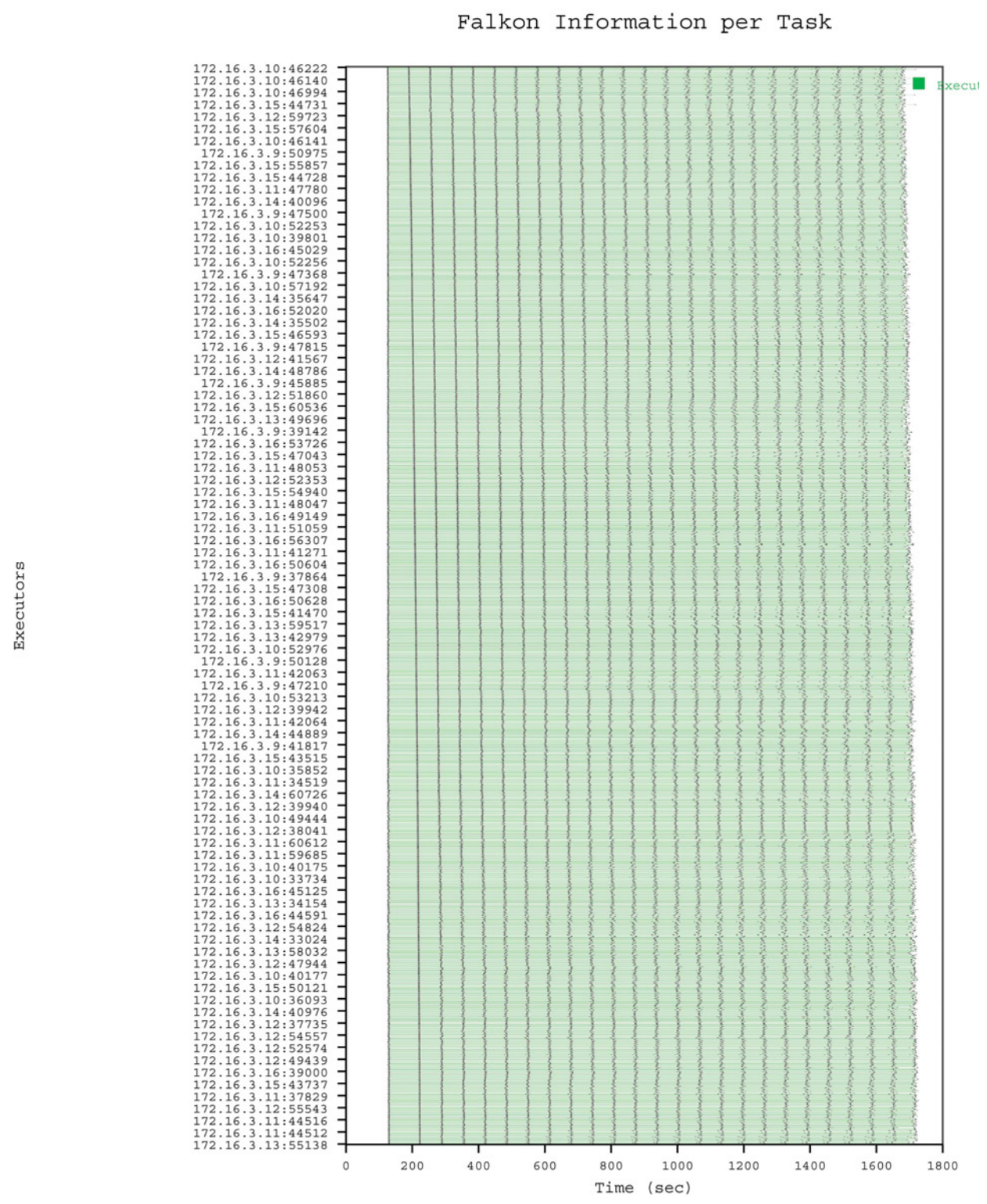

**Figure 18: MARS application (per processor view) on the BG/P; 7M micro-tasks (49K tasks) using 2048 processor cores**

The results presented in these figures are from a static workload processed directly with Falkon. Swift on the other hand can be used to make the workload more dynamic, reliable, and provide a natural flow from the results of this application to the input of the following stage in a more complex workflow. Swift incurs its own overheads in addition to what Falkon experiences when running the MARS application. These overheads include 1) managing the data (staging data in and out, copying data from its original location to a workflow-specific location, and back from the workflow directory to the result archival location) , 2) creating per-task working directories (via mkdir on the shared file system), and 3) creation and tracking of status logs files for each task.

We also ran a 16K task (2.4M micro-tasks) workload on 2048 CPUs which took an end-to-end time of 739.8 seconds. The per-micro-task time was higher than before – 0.602 seconds (up from 0.454 seconds in the 1 node/4 CPU case without any Swift or Falkon overhead). The efficiency between the average time per micro-task is 75%, but due to the slower dispatch rates (about 100 tasks/sec) and the higher variability in execution times (which yielded a slow ramp down of the experiment), the end-to-end efficiency was only 70% with a speedup of 1434X (2048 being ideal). The extra overhead (70% vs. 97% efficiency) between the Swift+Falkon execution and Falkon only execution can be attributed to the three things mentioned earlier (managing data, creating sand-boxes, and keeping track of status files, all on a per task basis).

It is interesting to note that Swift with the default settings and implementation, yielded only 20% efficiency for this workload. We investigated the main bottlenecks, and they seemed to be shared file system related. We applied three distinct optimizations to the Swift wrapper script: 1) the placement of temporary directories in local ramdisk rather than the shared filesystem; 2) copies the input data to the local ramdisk of the compute node for each job execution; and 3) creates the per job logs on local ramdisk and only copies them at the completion of each job (rather than appending a file on shared file system at each job status change). These optimizations allowed us to increase the efficiency from 20% to 70% on 2048 processors for the MARS application with task durations of 65.4 seconds (in ideal case).

We will be working to narrow the gap between the efficiencies found when running Swift and those when running Falkon alone, and hope to get the Swift efficiencies up in the 90% range without increasing the minimum task duration times per task. A relatively straight forward approach to increasing efficiency would be to increase the per task execution times, which could amortize the per task overhead better. However, at this stage of its development, 70% efficiency for a generic parallel scripting system running on 2K+ cores with 65 second tasks is a reasonable level of success.

# 6. Conclusions and Future Work

This paper focused on the ability to manage and execute large scale applications on petascale class systems. Clusters with 50K+ processor cores are beginning to come online (i.e. TACC Sun Constellation System - Ranger), Grids (i.e. TeraGrid) with a dozen sites and 100K+ processors, and supercomputers with 160K processors (i.e. IBM BlueGene/P). Large clusters and supercomputers have traditionally been high performance computing (HPC) systems, as they are efficient at executing tightly coupled parallel jobs within a particular machine with low-latency interconnects; the applications typically use message passing interface (MPI) to achieve the needed inter-process communication. On the other hand, Grids have been the preferred platform for more loosely coupled applications that tend to be managed and executed through workflow systems. In contrast to HPC (tightly coupled applications), the loosely coupled applications are known to make up high throughput computing (HTC). HTC systems generally involve the execution of independent, sequential jobs that can be individually scheduled on many different computing resources across multiple administrative boundaries. HTC systems achieve this using various grid computing techniques, and often times use files to achieve the inter-process communication (as opposed to MPI for HPC).

Our work shows that today's existing HPC systems are a viable platform to host loosely coupled HTC applications. We identified challenges that arise in large scale loosely coupled applications when run on petascale-precursor systems, which can hamper the efficiency and utilization of these large scale systems. These challenges vary from local resource manager scalability and granularity, efficient utilization of the raw hardware, shared file system contention and scalability, reliability at scale, application scalability, and understanding the limitations of the HPC systems in order to identify promising and scientifically valuable loosely-coupled applications. This paper presented new research, implementations, and applications experience in scaling loosely coupled large-scale applications on the IBM BlueGene/P and the SiCortex. Although our experiments are still on precursor systems (4K processors for the BG/P and 5.8K processors for the SiCortex), the experience we gathered is invaluable in planning to scale these applications another one to two orders of magnitude over the course of the next few months as the 160K processor BG/P comes online. We expect to present results on 40K-160K core systems in the final version of this paper.

For future work, we plan to implement and evaluate enhancements, such as task pre-fetching, alternative technologies, improved data management, and a three-tier architecture. Task pre-fetching is commonly done in manager-worker systems, where executors can request new tasks before they complete execution of old tasks, thus overlapping communication and execution. Many Swift applications read and write large amounts of data. Our efforts will in large part be focused on having all data management operations avoid the use of shared filesystem resources when local file-systems can handle the scale of data involved.

As we have seen in the results of this paper, data access is the main bottleneck as applications scale. We expect that data caching, proactive data replication, and data-aware scheduling will offer significant performance improvements for applications that exhibit locality in their data access patterns. [26] We have already implemented a data-aware scheduler, and support for caching in the Falkon Java executor. In previous work, we have shown that in both micro-benchmarks and a large-scale astronomy application, that a modest small Linux cluster (128 CPUs) can achieve aggregate I/O data rates of tens of Gb/s of I/O throughput [25, 27]. We plan to port the same data caching mechanisms from the Java executor to the C executor so we can use these techniques on the BG/P. Finally, we plan on evolving the Falkon architecture from the current 2-Tier architecture to a 3-Tier one. We are expecting that this architecture change will allow us to introduce more parallelism and distribution of the currently centralized management component in Falkon, and hence offer higher dispatch and execution rates than Falkon currently supports, which will be critical as we scale to the entire 160K-core BG/P and we get data caching implemented and running efficiently to avoid the shared file system overheads.